\documentclass[ aps, showkeys, nofootinbib, floatfix, superscriptaddress]{revtex4}

\usepackage{amsfonts}

\usepackage{amssymb}
\usepackage{amsmath}
\usepackage{graphicx}

\begin{document}

\title{From GR to STG --- Inheritance and development of Einstein's heritages}
 \author{Jianbo Lu}
 \email{lvjianbo819@163.com}
 \affiliation{Department of Physics, Liaoning Normal University, Dalian 116029, P. R. China}
  \author{Yongxin Guo}
 \email{guoyongxin@lnu.edu.cn}
 \affiliation{Department of Physics, Liaoning University, Shenyang, P. R. China}
 \author{G. Y. Chee}
 \email{qgy8475@sina.com}
 \affiliation{Department of Physics, Liaoning Normal University, Dalian 116029, P. R. China}

\begin{abstract}
A review of General Relativity, Teleparallel Gravity, and Symmetric Teleparallel gravity is given in this paper. By comparing these theories some conclusions are obtained. It is argued that the essence of gravity is the translation connection. The gauge group associated with gravity is merely the
translation group. Lorentz group is relevant with only inertial effects and has nothing to do with gravity. The Lorentz connection represents a inertial
force rather than a gauge potential, which leads to the equivalent principle. Neither curvature nor torsion belongs to spacetime but certain
transformation group and their connections. The statement "gravity is equivalent to curved spacetime" is a misunderstanding, not Einstein's real
intention. Not curvature but torsion i.e. the anholonomy of the tetrad represents gravity. Einstein's great genius is choosing metric as the
fundamental variable to describe gravity. Some conceptual issues, for example, general covariance, Lorentz covariance, reference system
dependence, gravitational energy-momentum are discussed.

\end{abstract}



\maketitle

\section{$\text{Introduction}$}

 General Relativity (GR) sets a precedent introducing geometry into physics. Einstein envisioned that the equivalence principle dictates the universality
and the geometrical nature of gravity. In GR the geometrization of the gravitational interaction is performed in terms of the metric and the
curvature. For a long time, gravity has been identified with the curvature, the gravitational phenomena have been recognized as a manifestation of a
curved spacetime [1]. A key point, however, is that the metric tensor can not define the curvature by itself generally and a connection is required
[2, 3]. In fact, there is no such thing as curvature of spacetime, but only curvature of connections. It seems far reasonable and convenient to take
spacetime simply as a manifold, and connections as additional structures. As an additional structure to the spacetime manifold, the connection can have
either vanishing or nonvanishing curvature and torsion. Consequently, various geometries and gravity theories are developed, for example, GR,
teleparallel gravity (TG) and symmetric teleparallel gravity (STG).

In GR the Christoffel connection plays a important role, which has a curvature but no torsion. On the contrary, in TG the connection has no curvature but
a torsion. Furthermore, there exists a gravity theory, the so called symmetric teleparallel gravity (STG), where the connection has neither the
curvature nor the torsion.

In addition to the connection the spacetime can be endowed with another geometry structure, a metric field on it. In GR and STG it is the
metric rather than the curvature that describes gravity. Furthermore, a more natural and intrinsic construction always present in any differentiable
manifold: at each point of the manifold there is a tangent space, a vector space attached to it. The bases (\emph{frames})\ of the \emph{ tangent space }are the very fundamental geometry objects and the most important variables describing gravity.

Now the problem is what role do the connection, the metric and the frame play in gravitational theories, respectively?

In order to unify with other interactions, gravity is expected to be formulated as a gauge theory. In the standard gauge theories the connection
of a principal bundle and its curvature are the fundamental variables. The connection representing the potential of the field is \emph{unobservable and}
can not be directly \emph{measured, }while the curvature representing field strength is \emph{observable and }can be directly \emph{measured}. Under the
corresponding gauge transformation the connection is not covariant and then can be transformed to zero. On the other hand, the curvature of the
connection is covariant and can never be transformed to zero.

Whether gravity can be described in terms of some principal bundle, for example, a frame bundle or the other bundle?

As a principal bundle the frame bundle is a constitutive part of spacetime. It is always present as soon as spacetime $M$, the base space of
the bundle, is considered as a differentiable manifold. The fiber over a point $x\in M$ is the set of all frames (ordered bases) for the tangent
space $T_{x}M$. In the most general case, the structure group $G$ of the bundle is the general \emph{linear group} $GL(4,\mathbb{R})$.

Once spacetime $M$ is endowed with a Lorentzian metric $g$, a \emph{sub-bundle} i.e. the \emph{bundle of orthonormal frames }can be defined. It
consists of orthonormal bases with respect to $g$, and with the \emph{Lorentz group} $SO(1,3)$ as the structure group [4]. The orthonormality
condition for the frames is introduced with the help of the soldering form [5].

The frame bundle is a principal bundle associated to the tangent bundle $M\times R_{4}$. The fiber of the latter is the tangent space spanned by the
tangent vectors at a point of the spacetime $M$. The tangent bundle $M\times R_{4}$ is locally homeomorphic to another bundle $M\times T_{4}$ of which
the fiber over a point $x\in M$ is a point set of the translation group $T_{4}$. This point set is an affine generalization of Minkowski [6], in
which the origin is not fixed [2]. The structure group of this bundle is the translation group $T_{4}$.

In recent years an important development is the discovery of separating inertial effects from gravity, and the existence of a true gravitational
variable in the usual sense of classical field theory [7, 8]. If gravity is a physically veritable force it must be represented by a covariant variable,
a curvature rather than a noncovariant connection. On the other hand, the inertial force as a fictitious forces should be represented by a
noncovariant variable, a connection for example, and can be transformed to zero by a gauge transformation.

It was proposed that the affine connection of a frame bundle should be associated with purely fictitious forces [9]. Except general relativity, in
all other relativistic theories the frame bundle connection has to do with inertial effects only [10].

In GR the connection plays the role of the field strength of gravitation rather than potential [11]. At the same time the potential is represent by
the metric tensor, a covariant variable. It is this concept conflict that makes GR can never be formulated as a gauge field theory.

The role played by the Lorentz connection seems to be a constant source of serious misunderstandings in the literature. In fact it has nothing to do
with gravitation although is behind all relativistic theories. It describes inertial effects only and does not have its own dynamic equation.\ Its role
is just to render all relativistic theories invariant under local Lorentz transformations [10]. The Lorentz connection is an alien object in the fiber
bundle formulation of the gravitational gauge theory.

In addition to the frame bundle there exists another fiber bundle, of which the structure group is the translation group $T_{4}$. Teleparallel gravity
(TG), a gauge theory for the translation group, is built on this geometrical structure. Compelling arguments to regard gravity as the gauge theory of
translations were presented e.g. by Feynman [12]. It can be consistently formulated in terms of the translational connection on a principal bundle
[13]. Gauge transformations are the local translations. The translational connection as the gravitational potential appears in the non-trivial part of
the tetrad field. The curvature\ is assumed to vanish from the very beginning, while the torsion is the gauge field strength and is interpreted
as the external curvature [14]. This reflects the special property of the gravitational interaction whose \textquotedblright
external\textquotedblright\ gauge geometry describes the spacetime itself [15]. Why a gauge theory for the translation group rather than any other
group is associated to spacetime? The answer is related to the source of gravitation -- energy and momentum. According to Noether's theorem, the
energy-momentum tensor is conserved provided the source Lagrangian is invariant under spacetime translations [16]. If gravity is described by a
gauge theory with energy-momentum as source, it must be a gauge theory for the translation group [13].

In TG inertial effects are separated from gravity successfully. The gravitational field is fully represented by the translational connection,
while the Lorentz connection keeps its special-relativistic role of representing the inertial effect only [6]. The gravitational force is
represent by a genuine strength of the gauge potential i.e. the torsion of the translation connection rather than the connection it self. The
fundamental field describing gravity is neither the frame nor the metric,
but the translational connection.

Although different as appears, TG is found to be equivalent to the description of GR. In contrast to GR, TG is a true field theory. In the
ordinary TG, we can obtain a number of fundamental insights into the nature of gravity, which are not readily available (or are, at least, more hidden)
in standard GR.

Most importantly, teleparallel gravity seems to be a much more appropriate theory to deal with the quantization of the gravitational field [5]. We can
then say that this theory is not just equivalent to general relativity, but a new way to look at all gravitational phenomena.

In addition to the ordinary teleparallel gravity there is a much less explored theory introduced by\ Nester and Yo [17]. In this theory a flat
connection is not metric compatible, and then the dynamics of GR can be reproduced in a teleparallel geometry even without torsion. Since the
teleparallel condition is satisfied and the connection is symmetric, this approach was named symmetric teleparallel gravity\ (STG).

In STG gravity is ascribed to the non-metricity and materialized in a flat and torsion free geometry. This framework leads to a purely inertial
connection [15] that can thus be completely removed by a coordinate gauge choice, the coincident gauge. Based on this property, it proposes a simpler
geometrical formulation of GR that is oblivious to the affine spacetime structure, thus fundamentally depriving gravity of any inertial character
[18]. Technically, calculations in coordinates are \textquotedblright legitimized\textquotedblright\ by bestowing covariance on them [17, 18]. The
geometrical framework for this formulation of GR is the simplest among the three equivalent representations.

STG brings a new perspective to bear on GR, which can uncover the more fundamental nature of the gravitational interaction. It keeps the original
Lagrangian and field equation of Einstein but abandon the curvature. It perhaps is closer to Einstein's own view of GR than the conventional
\textquotedblleft Einstein's GR\textquotedblright\ [19] and follows the
Principle Occam's Razor.

Although understood as a gauge theory for translations, STG has not been formulated by gauging translation in addition to the general linear group
but from within that group [19]. The theory is described by the metric rather than the translation connection or frame [14, 18]. It is also built
as a locally gauge invariant model [20], in which the\ metric represents the gravitational field while the connection corresponds to the gauge potential.
In another formulation, the gravitational potentials are represented by a cobase field [21].

\section{Geometry}

\subsection{Affine geometry Affine structure and connection}

Let $\left\{ b_{K}\right\} $ and $\left\{ b_{K}^{\prime }\right\} $ are two bases (\textbf{frames}) of a vector space. Any vector $V$ can be expressed as%
\begin{equation}
V=V^{K}b_{K}=V^{\prime K}b_{K}^{\prime }.
\end{equation}%
Under the base transformation

\begin{equation}
b_{K}^{\prime }=\Lambda _{K}{}^{L}b_{L},
\end{equation}%
the components $V^{K}$ of a vector transform as
\begin{equation}
V^{\prime K}=V^{L}\left( \Lambda ^{-1}\right) _{L}{}^{K},
\end{equation}%
where $\Lambda ^{-1}$ is the inverse of the matrix $\Lambda $.

In the case when $V$ is a vector field on the spacetime manifold $M$, i. e. $ V=V\left( x\right) ,x\in M,$ both $V^{K}=V^{K}\left( x\right) $ and $\Lambda
_{K}{}^{L}=\Lambda _{K}{}^{L}\left( x\right) $ are functions of the coordinates $x^{\mu }$ of $x$. Under the transformation (2), the derivative
transforms also:
\begin{equation}
\partial _{\mu }V^{\prime K}=\partial _{\mu }V^{L}\left( \Lambda
^{-1}\right) _{L}{}^{K\prime }+V^{L}\partial _{\mu }\left( \Lambda
^{-1}\right) _{L}{}^{K\prime }.
\end{equation}%
It is not covariant with (3) due to the second term. We introduce an affine connection $\Gamma ^{K}{}_{L\mu }=\Gamma ^{K}{}_{L\mu }\left( x\right) $ and
define the covariant derivative
\begin{equation}
D_{\mu }V^{K}:=\partial _{\mu }V^{K}+\Gamma ^{K}{}_{L\mu }V^{L}.
\end{equation}
\ \ \ \ \ \ \ \ \ \ \ Then we have
\begin{eqnarray*}
&&\partial _{\mu }V^{\prime K}+\Gamma ^{\prime K}{}_{L\mu }V^{\prime L} \\
&=&\left( \partial _{\mu }V^{L}+\Gamma ^{L}{}_{J\mu }V^{J}\right) \left(
\Lambda ^{-1}\right) _{L}{}^{K}+V^{J}\left\{ \partial _{\mu }\left( \Lambda
^{-1}\right) _{J}{}^{K}-\Gamma ^{L}{}_{J\mu }\left( \Lambda ^{-1}\right)
_{L}{}^{K}+\Gamma ^{\prime K}{}_{L\mu }\left( \Lambda ^{-1}\right)
_{J}{}^{L}\right\} ,
\end{eqnarray*}%
Letting%
\begin{equation*}
\partial _{\mu }\left( \Lambda ^{-1}\right) _{J}{}^{K}-\Gamma ^{L}{}_{J\mu
}\left( \Lambda ^{-1}\right) _{L}{}^{K}+\Gamma ^{\prime K}{}_{L\mu }\left(
\Lambda ^{-1}\right) _{J}{}^{L}=0,
\end{equation*}
we have%
\begin{equation*}
\Gamma ^{\prime K}{}_{I\mu }=\Lambda _{I}{}^{J}\Gamma ^{L}{}_{J\mu }\left(
\Lambda ^{-1}\right) _{L}{}^{K}-\Lambda _{I}{}^{J}\partial _{\mu }\left(
\Lambda ^{-1}\right) _{J}{}^{K}
\end{equation*}
or%
\begin{equation}
\Gamma ^{\prime K}{}_{I\mu }=\left( \Lambda ^{-1}\right) _{L}{}^{K}\Gamma
^{L}{}_{J\mu }\Lambda _{I}{}^{J}+\left( \Lambda ^{-1}\right)
_{J}{}^{K}\partial _{\mu }\Lambda _{I}{}^{J},
\end{equation}
and then%
\begin{equation*}
\partial _{\mu }V^{\prime K}+\Gamma ^{\prime K}{}_{L\mu }V^{\prime L}=\left(
\partial _{\mu }V^{L}+\Gamma ^{L}{}_{J\mu }V^{J}\right) \left( \Lambda
^{-1}\right) _{L}{}^{K},
\end{equation*}
i.e.%
\begin{equation}
D_{\mu }V^{\prime K}=D_{\mu }V^{L}\left( \Lambda ^{-1}\right) _{L}{}^{K}.
\end{equation}%
(7) is covariant with (3) which is the reason for name of the "covariant"
derivative $D_{\mu }$. However, (6) means that the connection $\Gamma
^{K}{}_{L\mu }$ is not covariant due to the second therm in (6).

The tangent space $T_{x}M$ of the spacetime manifold $M$ at any point $x\in M
$, is a vector space which is a fiber of the tangent bundle. As differential
operators the base vectors $b_{K}=b_{K}{}^{\mu }\partial _{\mu }$ satisfy
the commutation rule [1]
\begin{equation}
\lbrack b_{K},b_{L}]=f^{I}{}_{KL}b_{I},
\end{equation}%
where%
\begin{equation}
f^{I}{}_{KL}=b_{K}{}^{\mu }b_{L}{}^{\nu }\left( \partial _{\nu }b^{I}{}_{\mu
}-\partial _{\mu }b^{I}{}_{\nu }\right) ,
\end{equation}%
and $b^{I}{}=b^{I}{}_{\mu }dx^{\mu }$ is the coframe 1-form dual to $b_{K}$. If $f^{I}{}_{KL}=0,$the basis $\left\{ b_{K}\right\} $ is holonomic and if $
f^{I}{}_{KL}\neq 0$, $\left\{ b_{K}\right\} $ is aholonomic.

At any point $x\in M$ with the coordinate system $\left\{ x^{\mu }\right\} $, the coordinate basis $\left\{ \partial _{\mu }\right\} $ spans the tangent
space $T_{x}M$, which has its own coordinate system $\left\{ x^{a}\right\} $ and the basis $\left\{ \partial _{a}\right\} $. Usually, $\left\{ x^{\mu
}\right\} $ is called the external (world) coordinate, while $\left\{
x^{a}\right\} $ is called the internal coordinate. The basis $\left\{
\partial _{\mu }\right\} $ is holonomic while the basis $\left\{ \partial
_{a}\right\} $ is aholonomic. Then we have three kinds of coordinate transformations:

i) The internal coordinate transformations $x^{a}\rightarrow x^{\prime a}$, the coordinate base vectors $\partial _{a}$ transform as%
\begin{equation*}
\partial _{a}\longrightarrow \partial _{a}^{\prime }=\frac{\partial x^{b}}{%
\partial x^{\prime a}}\partial _{b}=L_{a}{}^{b}\partial _{b}.
\end{equation*}%
By the substitutions%
\begin{equation*}
\Lambda _{K}{}^{L}\rightarrow L_{a}{}^{b},\Gamma ^{L}{}_{J\mu }\rightarrow
A^{a}{}_{b\mu }
\end{equation*}
(6) takes the form
\begin{equation}
A^{\prime a}{}_{b\mu }=\left( L^{-1}\right) _{c}{}^{a}A^{c}{}_{d\mu
}L_{b}{}^{d}+\left( L^{-1}\right) _{c}{}^{a}\partial _{\mu }L_{b}{}^{c}.
\end{equation}

ii) The external (world) coordinate transformations $x^{\mu }\rightarrow
x^{\prime \mu }$, the coordinate base vectors $\partial _{\mu }$ transform as
\begin{equation*}
\partial _{\mu }\longrightarrow \partial _{\mu }^{\prime }=\frac{\partial
x^{\nu }}{\partial x^{\prime \mu }}\partial _{\nu }=M_{\mu }{}^{\nu
}\partial _{\nu }.
\end{equation*}
By the substitutions
\begin{equation*}
\Lambda _{K}{}^{L}\rightarrow \frac{\partial x^{\prime \nu }}{\partial
x^{\mu }},\left( \Lambda ^{-1}\right) _{L}{}^{K}\rightarrow \frac{\partial
x^{\nu }}{\partial x^{\prime \sigma }},\Gamma ^{L}{}_{J\mu }\rightarrow
\Gamma ^{\rho }{}_{\sigma \mu },
\end{equation*}
(6) takes the form
\begin{equation}
\Gamma ^{\prime \lambda }{}_{\tau \mu }=\frac{\partial x^{\lambda }}{
\partial x^{\prime \rho }}\frac{\partial x^{\nu }}{\partial x^{\prime \mu }}
\frac{\partial x^{\prime \sigma }}{\partial x^{\tau }}\Gamma ^{\rho
}{}_{\sigma \nu }+\frac{\partial x^{\lambda }}{\partial x^{\prime \rho }}
\frac{\partial x^{\nu }}{\partial x^{\prime \mu }}\frac{\partial
^{2}x^{\prime \rho }}{\partial x^{\nu }\partial x^{\tau }}.
\end{equation}

iii) The transformations between the external and the internal coordinates, and their base vector
\begin{equation}
x^{\mu }=x^{\mu }\left( x^{a}\right) ,\partial _{\mu }\rightarrow \partial
_{a}=\frac{\partial x^{\mu }}{\partial x^{a}}\partial _{\mu }=e_{a}{}^{\mu
}\partial _{\mu },
\end{equation}%
and vice versa

\begin{equation}
x^{a}=x^{a}\left( x^{\mu }\right) ,\partial _{a}\rightarrow \partial _{\mu }=
\frac{\partial x^{a}}{\partial x^{\mu }}\partial _{a}=e^{a}{}_{\mu }\partial
_{a}.
\end{equation}
By the substitutions
\begin{eqnarray*}
b_{K} &\rightarrow &\partial _{a},b_{K}^{\prime }\rightarrow \partial _{\mu
},V^{K}\rightarrow e^{a} \\
\left( \Lambda ^{-1}\right) _{L}{}^{K} &\rightarrow &e_{a}{}^{\rho },\Lambda
_{I}{}^{J}\rightarrow e^{b}{}_{\sigma } \\
\Gamma ^{L}{}_{J\mu } &\rightarrow &A^{a}{}_{b\mu },\Gamma ^{\prime
K}{}_{I\mu }\rightarrow \Gamma ^{\rho }{}_{\mu \nu }
\end{eqnarray*}
(5) and (6) take the forms
\begin{equation}
D_{\mu }e^{a}{}_{\nu }=\partial _{\mu }e^{a}{}_{\nu }+A^{a}{}_{b\mu
}e^{b}{}_{\nu },
\end{equation}
and%
\begin{equation}
\Gamma ^{\rho }{}_{\mu \nu }=e_{a}{}^{\rho }\partial _{\mu }e^{a}{}_{\nu
}+e_{a}{}^{\rho }A^{a}{}_{b\mu }e^{b}{}_{\nu }\equiv e_{a}{}^{\rho }D_{\mu
}e^{a}{}{}_{\nu }.
\end{equation}
The inverse relation is, consequently,
\begin{equation}
\nabla _{\mu }e_{b}{}^{\nu }=\partial _{\mu }e_{b}{}^{\nu }+\Gamma ^{\nu
}{}_{\mu \rho }e_{b}{}^{\rho },
\end{equation}
\begin{equation}
A^{a}{}_{b\mu }=e^{a}{}_{\nu }\partial _{\mu }e_{b}{}^{\nu }+e^{a}{}_{\rho
}\Gamma ^{\rho }{}_{\mu \nu }e_{b}{}^{\nu }\equiv e^{a}{}_{\nu }\nabla _{\mu
}e_{b}{}^{\nu },
\end{equation}
where $\nabla _{\mu }$ is the covariant derivative with respect to the connection $\Gamma ^{\rho }{}_{\nu \mu }$, which acts on external indices $
\mu ,\nu ,...$\ only, while $D_{\mu }$ is the covariant derivative with respect to the connection $A^{a}{}_{b\mu }$ which acts on internal (or
tangent space) indices $a,b,...$\ only. For example, The actions of $\nabla _{\mu }$ on vectors $V^{\alpha }$ and 1-forms $V_{\alpha }$ are,
respectively,
\begin{eqnarray}
\nabla _{\mu }V^{\alpha } &=&V^{\alpha }{}_{,\mu }+\Gamma ^{\alpha }{}_{\mu
\lambda }V^{\lambda },\;\;\;\;\;\;\;\; \\
\nabla _{\mu }V_{\alpha } &=&V{}_{\alpha ,\mu }-\Gamma ^{\lambda }{}_{\mu
\alpha }V_{\lambda }.\;\;\;\;\;\;\;\;
\end{eqnarray}
Then \emph{both }$\Gamma ^{\alpha }{}_{\mu \nu }$\emph{\ and }$A^{a}{}_{b\mu }$\emph{\ represent the one and the same connection in different disguises}.
This conclusion is further substantiated by comparing the torsions and the curvatures for both connections,%
\begin{equation}
T^{a}{}_{\mu \nu }\left( A\right) =\partial _{\mu }e^{a}{}_{\nu }-\partial
_{\nu }e^{a}{}_{\mu }+A^{a}{}_{c\mu }e^{c}{}_{\nu }-A^{a}{}_{c\nu
}e^{c}{}_{b\mu },
\end{equation}

\begin{equation}
R^{a}{}_{b\mu \nu }\left( A\right) =\partial _{\mu }A^{a}{}_{b\nu }-\partial
_{\nu }A^{a}{}_{b\mu }+A^{a}{}_{c\mu }A^{c}{}_{b\nu }-A^{a}{}_{c\nu
}A^{c}{}_{b\mu },
\end{equation}
and%
\begin{equation}
T^{\rho }{}_{\mu \nu }\left( \Gamma \right) =\Gamma ^{\rho }{}_{\mu \nu
}-\Gamma ^{\rho }{}_{\mu \nu },
\end{equation}

\begin{equation}
R^{\alpha }{}_{\beta \mu \nu }\left( \Gamma \right) =\partial _{\mu }\Gamma
^{\alpha }{}_{\beta \nu }-\partial _{\nu }\Gamma ^{\alpha }{}_{\mu \beta
}+\Gamma ^{\alpha }{}_{\rho \mu }\Gamma ^{\rho }{}_{\beta \nu }-\Gamma
^{\alpha }{}_{\rho \nu }\Gamma ^{\rho }{}_{\beta \mu },\;
\end{equation}
which after a simple calculation gives

\begin{equation}
T^{\rho }{}_{\mu \nu }\left( \Gamma \right) =e_{a}{}^{\rho }T^{a}{}_{\mu \nu
}\left( A\right) ,
\end{equation}

\begin{equation}
R^{\alpha }{}_{\beta \mu \nu }\left( \Gamma \right) =e_{a}{}^{\alpha }e^{b}{}_{\beta }R^{a}{}_{b\mu \nu }\left( A\right) .
\end{equation}

The connection $A^{a}{}_{b\mu }$\ is independent of the frame (tetrad) $ e^{a}{}_{\mu }$. The torsion tensor is a function of both the tetrad $
e^{a}{}_{\mu }$ and the connection $A^{a}{}_{b\mu }$.

For the internal coordinate $x^{a}$, in addition to the basis transformations there is another kind of transformations, the translation on
$T_{x}M,$
\begin{equation}
x^{a}\rightarrow x^{\prime a}=x^{a}+\varepsilon ^{a}\left( x^{\mu }\right) .
\end{equation}
Under these transformations a vector $V^{a}$ transform as
\begin{equation}
V^{a}\rightarrow V^{\prime a}=V^{a},
\end{equation}
while the derivative $\partial _{\mu }V^{a}$, transforms according to the
rule
\begin{equation}
\partial _{\mu }V^{\prime a}=\partial _{\mu }V^{a}+\partial _{\mu
}\varepsilon ^{b}\partial _{b}V^{a}.
\end{equation}
We introduce the translation connection $B^{a{}}{}_{\mu }$ and covariant derivative
\begin{equation}
h_{\mu }^{\left( 0\right) }V^{a}=\left( \partial _{\mu }x^{b}+B^{b{}}{}_{\mu
}\right) \partial _{b}V^{a},
\end{equation}
and compute
\begin{equation*}
\partial _{\mu }V^{\prime a}+B^{\prime b{}}{}_{\mu }\partial _{b}V^{\prime
a}=\partial _{\mu }V^{a}+B^{b{}}{}_{\mu }\partial _{b}V^{a}+\left( B^{\prime
b{}}{}_{\mu }-B^{b{}}{}_{\mu }+\partial _{\mu }\varepsilon ^{b}\right)
\partial _{b}V^{a}.
\end{equation*}

If
\begin{equation}
B^{\prime b{}}{}_{\mu }=B^{b{}}{}_{\mu }-\partial _{\mu }\varepsilon ^{b},
\end{equation}
\ \ \ \ we have
\begin{equation*}
\partial _{\mu }x^{\prime b}\partial _{b}V^{\prime a}+B^{\prime b{}}{}_{\mu
}\partial _{b}V^{\prime a}=\partial _{\mu }x^{b}\partial
_{b}V^{a}+B^{b{}}{}_{\mu }\partial _{b}V^{a},
\end{equation*}
i.e.%
\begin{equation}
h_{\mu }^{\left( 0\right) }V^{\prime a}=h_{\mu }^{\left( 0\right) }V^{a},
\end{equation}
which means that $h_{\mu }^{\left( 0\right) }V^{a}$ is covariant with $V^{a}$.

An obvious but forgot often fact is that any connection is not covariant under the corresponding transformations. Here we emphasize "the
corresponding". As an example, $A^{c}{}_{b\mu }$ is not covariant under the internal coordinate transformations $L^{a}{}_{b}$, due to the term $
L^{a}{}_{b}\partial _{\mu }\left( L^{-1}\right) ^{b}{}_{d}$. However, as a 1-form, it is covariant under the external coordinate transformations.
Similarly, $\Gamma ^{\sigma }{}_{\rho \tau }$ is not covariant under the external coordinate transformations, but is invariant under the internal
coordinate transformations as a scalar. By the same way, the term $\partial
_{\mu }\varepsilon ^{b}$ in (30) makes $B^{b{}}{}_{\mu }$ not covariant under translations $x^{\prime a}=x^{a}+\varepsilon ^{a}$. However, it being
a vector in internal coordinates and a 1-form in external coordinates is covariant under both transformations $L_{a}{}^{b}=\frac{\partial x^{b}}{
\partial x^{\prime a}}$ and $M_{\mu }{}^{\nu }=\frac{\partial x^{\nu }}{
\partial x^{\prime \mu }}$. The key point is to remember a geometry object can have different properties under different transformations.

Note that the Lorentz potential $A^{a}{}_{b\nu }$ is invariant under the local translations, while the translation potential $B^{a}{}_{\mu }$ (a
Lorentz vector and a coordinate 1-form) is covariant under the Lorentz transformations. The gauge potential $\Gamma ^{\nu }{}_{\mu \lambda }$ of
the coordinate transformation group is not covariant:
\begin{equation*}
\Gamma ^{\prime \nu }{}_{\mu \lambda }=\frac{\partial x^{\rho }}{\partial
x^{\prime \mu }}\left( \frac{\partial x^{\tau }}{\partial x^{\prime \lambda }
}\frac{\partial x^{\nu }}{\partial x^{\sigma }}\Gamma ^{\sigma }{}_{\rho
\tau }+\frac{\partial x^{\sigma }}{\partial x^{\prime \lambda }}\frac{
\partial ^{2}x^{\nu }}{\partial x^{\rho }\partial x^{\sigma }}\right) ,
\end{equation*}%
but is invariant under both the local translations and the local Lorentz
transformations.

The covariant derivative operator
\begin{equation}
h_{\mu }^{\left( 0\right) }=\left( \partial _{\mu }x^{b}+B^{b{}}{}_{\mu
}\right) \partial _{b}=h^{\left( 0\right) b{}}{}_{\mu }\partial _{b}\text{ \
\ \ \ }
\end{equation}
is a 1-form
\begin{equation}
h^{\left( 0\right) b{}}=h^{\left( 0\right) b{}}{}_{\mu }dx^{\mu },
\end{equation}
which is dual to the vector
\begin{equation}
h_{a}^{\left( 0\right) }=h_{a}^{\left( 0\right) }{}^{\mu }\partial _{\mu }.
\end{equation}
As differential operators, $h_{a}^{\left( 0\right) }$ satisfy the
commutation rule
\begin{equation}
\lbrack h_{a}^{\left( 0\right) },h_{b}^{\left( 0\right)
}]=f^{c}{}_{ab}h_{c}^{\left( 0\right) },
\end{equation}
where
\begin{equation}
f^{c}{}_{ab}=h_{a}^{\left( 0\right) }{}^{\mu }h_{b}^{\left( 0\right)
}{}^{\nu }(\partial _{\nu }h^{\left( 0\right) c}{}_{\mu }-\partial _{\mu
}h^{\left( 0\right) c}{}_{\nu }).
\end{equation}
Recalling
\begin{equation}
h^{\left( 0\right) b{}}{}_{\mu }=\partial _{\mu }x^{b}+B^{b{}}{}_{\mu }\text{
\ \ \ }
\end{equation}
we have%
\begin{equation}
f^{c}{}_{ab}=h_{a}^{\left( 0\right) }{}^{\mu }h_{b}^{\left( 0\right)
}{}^{\nu }(\partial _{\nu }B^{c}{}_{\mu }-\partial _{\mu }B^{c}{}_{\nu }).
\text{ \ }
\end{equation}
In the case $B^{b{}}{}_{\mu }=0$ we have
\begin{equation}
h^{\left( 0\right) b{}}{}_{\mu }=e^{b{}}{}_{\mu }=\partial _{\mu }x^{b},
\text{ \ \ }
\end{equation}
and%
\begin{equation}
f^{c}{}_{ab}=0.\text{ \ \ \ }
\end{equation}

We get the conclusion that the translation connection $B^{a}${}$_{\mu }$\
corresponds to an aholonomic basis (tetrad) $h^{\left( 0\right) a{}}${}$
_{\mu }${\LARGE .}

Considering the two transformations, the linear transformation and the translation, at the same time,
\begin{equation}
x^{a}\rightarrow x^{\prime a}=L^{a}{}_{b}x^{b}+\varepsilon ^{a},\text{ \ \ \
}
\end{equation}
we must take $D_{\mu }x^{b}$ instead of $\partial _{\mu }x^{b}$, and then have
\begin{eqnarray}
h_{\mu }V^{a} &=&\left( D_{\mu }x^{b}+B^{b{}}{}_{\mu }\right) \partial
_{b}V^{a}=\left( \partial _{\mu }x^{b}+A^{b}{}_{c\mu }x^{c}+B^{b{}}{}_{\mu
}\right) \partial _{b}V^{a}  \notag \\
&=&\left( h^{\left( 0\right) b{}}{}_{\mu }+A^{b}{}_{c\mu }x^{c}\right)
\partial _{b}V^{a}  \notag \\
&=&h^{b{}}{}_{\mu }\partial _{b}V^{a},\text{ \ \ \ \ \ \ \ \ \ \ \ \ \ \ \ \
\ \ \ \ \ \ \ \ \ \ \ \ \ \ \ }
\end{eqnarray}
where
\begin{equation}
h^{b{}}{}_{\mu }:=h^{\left( 0\right) b{}}{}_{\mu }+A^{b}{}_{c\mu
}x^{c}=\partial _{\mu }x^{b}+A^{b}{}_{c\mu }x^{c}+B^{b{}}{}_{\mu
}=e^{b{}}{}_{\mu }+A^{b}{}_{c\mu }x^{c}+B^{b{}}{}_{\mu }.\text{ \ \ }
\end{equation}

A local translation transforms trivial\ frame $e^{a}{}{}_{\mu }=\partial
_{\mu }x^{a}$\ to a nontrivial frame $h^{a}{}{}_{\mu }$\ and introduces a
gauge (gravitational force) potential $B^{a}{}_{\mu }$. A connection $
A^{b}{}_{c\mu }$ is always hidden in the tetrad $h^{b{}}{}_{\mu }$. It
appears when we pass to the general frame by performing a local frame
transformation $L_{a}{}^{b}$. Therefore, the general tetrad $h^{b{}}{}_{\mu
} $ is given by a combination of the connections $A^{b}{}_{c\mu }$ and $
B^{b{}}{}_{\mu }$.

The equations (15), (17) and (20) now become, respectively,
\begin{equation}
\Gamma ^{\rho }{}_{\mu \nu }=h_{a}{}^{\rho }\partial _{\mu }h^{a}{}_{\nu
}+h_{a}{}^{\rho }A^{a}{}_{b\mu }h^{b}{}_{\nu }\equiv h_{a}{}^{\rho }D_{\mu
}h^{a}{}{}_{\nu }.\text{ \ \ \ \ }
\end{equation}
\begin{equation}
A^{a}{}_{b\mu }=h^{a}{}_{\nu }\partial _{\mu }h_{b}{}^{\nu }+h^{a}{}_{\rho
}\Gamma ^{\rho }{}_{\nu \mu }h_{b}{}^{\nu }\equiv h^{a}{}_{\nu }\nabla _{\mu
}h_{b}{}^{\nu },\text{ \ \ \ }
\end{equation}
\begin{equation}
T^{a}{}_{\mu \nu }\left( A\right) =\partial _{\mu }h^{a}{}_{\nu }-\partial
_{\nu }h^{a}{}_{\mu }+A^{a}{}_{c\mu }h^{c}{}_{\nu }-A^{a}{}_{c\nu
}h^{c}{}_{\mu }.\text{ \ \ \ }
\end{equation}

As differential operators $h_{a}$ satisfy the commutation rule
\begin{equation}
\lbrack h_{a},h_{b}]=f^{c}{}_{ab}h_{c},\text{ \ \ \ \ }
\end{equation}
where%
\begin{equation}
f^{c}{}_{ab}=h_{a}{}^{\mu }h_{b}{}^{\nu }(\partial _{\nu }h^{c}{}_{\mu
}-\partial _{\mu }h^{c}{}_{\nu }).\text{ \ \ \ \ \ \ \ \ \ \ \ \ \ }
\end{equation}
Recalling (43) and (21) we have
\begin{equation}
f^{c}{}_{ab}=h_{a}{}^{\mu }h_{b}{}^{\nu }(\left\{ \partial _{\nu
}A^{c}{}_{d\mu }-\partial _{\mu }A^{c}{}_{d\nu }\right\} x^{d}+A^{c}{}_{d\mu
}\partial _{\nu }x^{d}-A^{c}{}_{d\nu }\partial _{\mu }x^{d}+\partial _{\nu
}B^{c{}}{}_{\mu }-\partial _{\mu }B^{c{}}{}_{\nu }),
\end{equation}
and%
\begin{eqnarray}
T^{a}{}_{\mu \nu }\left( A\right)  &=&R^{a}{}_{b\mu \nu }\left( A\right)
x^{b}\text{\ }+\partial _{\mu }B^{a{}}{}_{\nu }+A^{a}{}_{b\mu
}B^{b{}}{}_{\nu }-\partial _{\nu }B^{a{}}{}_{\mu }-A^{a}{}_{b\nu
}B^{b{}}{}_{\mu }  \notag \\
&=&R^{a}{}_{b\mu \nu }\left( A\right) x^{b}\text{\ }+D_{\mu }B^{a{}}{}_{\nu
}-D_{\nu }B^{a{}}{}_{\mu }.
\end{eqnarray}

In the case
\begin{equation*}
B^{b{}}{}_{\mu }=0,
\end{equation*}
(50) gives
\begin{equation*}
T^{a}{}_{\mu \nu }\left( A\right) =R^{a}{}_{b\mu \nu }\left( A\right) x^{b}.
\end{equation*}
In the gauge
\begin{equation*}
A^{a}{}_{b\mu }=0,
\end{equation*}
we have
\begin{equation*}
T^{a}{}_{\mu \nu }\left( A\right) =R^{a}{}_{b\mu \nu }\left( A\right) =0.
\end{equation*}
But if
\begin{equation*}
B^{b{}}{}_{\mu }\neq 0,\ A^{a}{}_{b\mu }=0,
\end{equation*}
we have
\begin{equation*}
T^{a}{}_{\mu \nu }=\partial _{\mu }h^{a}{}_{\nu }-\partial _{\nu
}h^{a}{}_{\mu }=\partial _{\mu }B^{a}{}_{\nu }-\partial _{\nu }B^{a}{}_{\mu
}=F^{a}{}_{\mu \nu }=h^{b}{}_{\mu }h^{a}{}_{\nu }f^{c}{}_{ab}.\text{ \ \ }
\end{equation*}
This means that $T^{a}{}_{\mu \nu }$ is the curvature of $B^{a{}}{}_{\mu }$
and the anholonomy of $h^{a{}}{}_{\mu }$.

Now, we have found an affine bundle with the spacetime as its base space.
This is a principal bundle with affine group as its structure group, which
is semidirect product of the general linear group $GL(4,\mathbb{R})$ and the
translation group $T_{4}$. The corresponding connections are $\Gamma
^{\alpha }{}_{\mu \nu }$($A^{a}{}_{b\mu }$) and $B^{a{}}{}_{\mu }$,respectively.

\subsection{Metric geometry}

A metric tensor $\mathbf{g}$ can be introduced on $M$, i.e. for the bases $ \left\{ \partial _{\mu }\right\} $ and $\left\{ h_{a}\right\} $,  and we have,
respectively,
\begin{equation}
\mathbf{g}\left( \partial _{\mu },\partial _{\nu }\right) =g_{\mu \nu
}\left( x\right),
\end{equation}
\begin{equation}
\mathbf{g}\left( h_{a},h_{b}\right) =\eta _{ab}\left( x\right),
\end{equation}
where
\begin{eqnarray}
g_{\mu \nu } &=&h^{a}{}_{\mu }h^{b}{}_{\nu }\eta _{ab}, \\
\eta _{ab}\left( x\right)  &=&h_{a}{}^{\mu }h_{b}{}^{\nu }g_{\mu \nu }.
\end{eqnarray}
If $\eta _{ab}=$ diag $\left( -1,1,1,1\right) $, \{$h_{a}$\} is called a Lorentz frame or a tetrad, the transformations \{$L^{a}{}_{b}$\} constitute
a Lorentz group $SO(1,3)$. Now we get a sub-bundle i.e. the bundle of orthonormal frames \{$h_{a}$\}. It consists of orthonormal bases with
respect to $g$, and with the Lorentz group $SO(1,3)$ as the structure group [4].

Being different from general linear transformations Lorentz transformations have specific physical meaning. They deal essentially with inertial effects.
Different classes of frames are obtained by performing local Lorentz transformations. Within each class, the infinitely many frames are related
to each other by global Lorentz transformations.

Einstein's idea is to identify gravity with geometry of the spacetime. The basic geometrical objects on the spacetime manifold are the sets of the
basis vectors, the frames. The frame $h^{b{}}{}_{\mu }:=\partial _{\mu
}x^{b}+A^{b}{}_{c\mu }x^{c}+B^{b{}}{}_{\mu }$ is endowed with abundant
physical meaning. The holonomic frame $e^{b{}}{}_{\mu }=\partial _{\mu
}x^{b} $ describes a inertial reference system. The anholonomic frame $
h^{\left( 0\right) b{}}{}_{\mu }=\partial _{\mu }x^{b}+B^{b{}}{}_{\mu }$ is called a proper frame representing a non-inertial reference system and
describes gravitation, while the anholonomic frame $h^{\left( L\right)
b{}}{}_{\mu }=\partial _{\mu }x^{b}+A^{b}{}_{c\mu }x^{c}$ is also a non-inertial reference system and describes the inertial force. $
B^{b{}}{}_{\mu }$ is the gravitational potential, while $A^{b}{}_{c\mu }x^{c}
$ corresponds the "potential\textquotedblright\ of the inertial force $ A^{b}{}_{c\mu }$.

Once a metric field is defined on the spacetime manifold, the non-metricity
\begin{equation}
Q_{\mu \alpha \beta }\equiv \nabla _{\mu }g_{\alpha \beta }=\partial _{\mu
}g_{\alpha \beta }-\Gamma ^{\rho }{}_{\alpha \mu }g_{\rho \beta }-\Gamma
^{\rho }{}_{\beta \mu }g_{\alpha \rho },
\end{equation}
arises as a part of the affine connection $\Gamma {}_{\mu \alpha \beta
}=g_{\mu \nu }\Gamma ^{\nu {}}{}_{\alpha \beta }$ which can be decomposed
into three parts:
\begin{equation}
\Gamma {}_{\mu \alpha \beta }=\left\{ _{\alpha \mu \beta }\right\} +K_{\mu
\alpha \beta }+L_{\mu \alpha \beta },
\end{equation}
where
\begin{eqnarray}
\left\{ _{\alpha \mu \beta }\right\}  &=&\frac{1}{2}\left( \partial _{\alpha
}g_{\mu \beta }+\partial _{\beta }g_{\alpha \mu }-\partial _{\mu }g_{\alpha
\beta }\right) ,\text{ } \\
K_{\mu \alpha \beta } &=&\frac{1}{2}\left( T{}_{\beta \alpha \mu }+T{}_{\mu
\alpha \beta }+T_{\alpha \beta \mu }\right) ,\text{ \ \ \ } \\
L_{\mu \alpha \beta } &=&\frac{1}{2}\left( \nabla _{\mu }g_{\alpha \beta
}-\nabla _{\alpha }g_{\mu \beta }-\nabla _{\beta }g_{\alpha \mu }\right) ,
\end{eqnarray}%
are referred as the Christoffel (or Levi-Civita) connection, the contortion and the disformation separately.

The components $\left\{ \Gamma ^{\rho }{}_{\mu \nu }\right\} $ and $\left\{
A^{a}{}_{b\mu }\right\} $ of the unique affine connection in different frames are tend to be considered as different objects according to the
frames and their transforming rules. Furthermore, they are endowed with different geometries and corresponding gravity theories. Usually there are
three cases, the so called geometrical trinity of gravity: GR for $T{}_{\alpha \beta \mu }=0,\nabla _{\mu }g_{\alpha \beta }=0$; TG for $%
R^{a}{}_{b\mu \nu }=0,\nabla _{\mu }g_{\alpha \beta }=0$; and STG for $R^{a}{}_{b\mu \nu }=0,T{}_{\alpha \beta \mu }=0$. They are discussed in the
next sections.

\section{General relativity (GR)}

In the case $T{}_{\alpha \beta \mu }=0,\nabla _{\mu }g_{\alpha \beta }=0$,
(56) and (23) give
\begin{equation}
\Gamma {}_{\mu \alpha \beta }=\left\{ _{\alpha \mu \beta }\right\} ,
\end{equation}
and
\begin{eqnarray}
R^{\alpha }{}_{\beta \mu \nu }\left( \Gamma \right)  &=&R^{\alpha }{}_{\beta
\mu \nu }\left( \left\{ {}\right\} \right)   \notag \\
&=&\partial _{\mu }\left\{ _{\beta }{}^{\alpha }{}_{\nu }\right\} -\partial
_{\nu }\left\{ _{\mu }{}^{\alpha }{}_{\beta }\right\} +\left\{ _{\rho
}{}^{\alpha }{}_{\mu }\right\} \left\{ _{\beta }{}^{\rho }{}_{\nu }\right\}
-\left\{ _{\rho }{}^{\alpha }{}_{\nu }\right\} \left\{ _{\beta }{}^{\rho
}{}_{\mu }\right\} .\;
\end{eqnarray}

In this case the connection $\Gamma {}_{\mu \alpha \beta }$ and the metric $ g_{\mu \nu }$ are not independent each other any more, but related by the
equation (60). The curvature $R^{\alpha }{}_{\beta \mu \nu }\left( \Gamma \right) $ becomes the curvature $R^{\alpha }{}_{\beta \mu \nu }\left(
\left\{ {}\right\} \right) $\ of the Chriatoffel connection $\left\{ _{\mu
}{}^{\alpha }{}_{\nu }\right\} $ and is called the curvature of the metric or the spacetime, which is the case of GR.

As the variable describing gravity we can only choose the metric $g_{\mu \nu
}$ or equivalently the frame $h^{a}{}_{\mu }$, which is related with the metric by $g_{\mu \nu }=\eta _{ab}h^{a}{}_{\mu }h^{b}{}_{\nu }$.

The frame $h^{b{}}{}_{\mu }=e^{b{}}{}_{\mu }+A^{b}{}_{c\mu
}x^{c}+B^{b{}}{}_{\mu }$ is covariant under local translations as well as local Lorentz transformations and so does the metric $g_{\mu \nu }$ $=\eta
_{ab}h^{a{}}{}_{\mu }h^{b{}}{}_{\nu }$. In GR the appearing of the connection $\Gamma ^{\lambda }{}_{\mu \nu }=\left\{ _{\mu \nu }^{\lambda
}\right\} $ in the covariant derivative guarantee the invariance of the theory under local translations, local Lorentz transformations and general
coordinate transformations. The frame $h^{b{}}{}_{\mu }=e^{b{}}{}_{\mu
}+A^{b}{}_{c\mu }x^{c}+B^{b{}}{}_{\mu }$ includes both the gravitation and the inertial effect and then so do the metric $g_{\mu \nu }$ $=\eta
_{ab}h^{a{}}{}_{\mu }h^{b{}}{}_{\nu }$ and the connection $\Gamma ^{\lambda
}{}_{\mu \nu }=\left\{ _{\mu \nu }^{\lambda }\right\} $. In other words, In GR, the inertial effect can not separate from the gravity.

Choosing the curvature scalar $R=g^{\beta \nu }R^{\alpha }{}_{\beta \alpha
\nu }$ as the gravitational Lagrangian leads to the Einstein field equation, which can be written as
\begin{equation}
R{}_{\mu \nu }-\frac{1}{2}Rg{}_{\mu \nu }=\left( \delta _{\mu }^{\rho
}\delta _{\nu }^{\sigma }-\frac{1}{2}g{}_{\mu \nu }g^{\rho \sigma }\right)
\nabla _{\lambda }\Gamma ^{\lambda }{}_{\rho \sigma }=T{}_{\mu \nu },
\end{equation}
where $T{}_{\mu \nu }$ is the energy-momentum tensor of the source. Comparing (62) with the gauge field equation
\begin{equation}
\nabla _{\nu }F_{i}{}^{\mu \nu }=J_{i}{}^{\mu },
\end{equation}%
and comparing the Einstein Lagrangian

\begin{equation}
\mathcal{L}=\sqrt{-g}g^{\mu \nu }\left( \left\{ _{\beta }{}^{\alpha }{}_{\mu
}\right\} \left\{ _{\nu }{}^{\beta }{}_{\alpha }\right\} -\left\{ _{\beta
}{}^{\alpha }{}_{\alpha }\right\} \left\{ _{\mu }{}^{\beta }{}_{\nu
}\right\} \right) ,
\end{equation}%
with the Yang- Mills Lagrangian

\begin{equation}
\mathcal{L}=\frac{1}{4}F_{i}{}^{\mu \nu }F^{i}{}_{\mu \nu },
\end{equation}
indicates that $\left\{ _{\beta }{}^{\alpha }{}_{\gamma }\right\} $ plays the role of the field strength $F_{i}{}^{\mu \nu }$, while $g{}_{\mu \nu }$
plays the role of the gauge potential as considered by Einstein originally [11]. However, $g{}_{\mu \nu }$ is not a connection but a quadratic function
of the translation connection $B^{a}{}_{\mu }$, as indicated by (53) and (43). The frame $h^{a}{}_{\mu }$ as well as the metric $g{}_{\mu \nu }$
include both the gravity and the inertial effect and so do the connection $\Gamma {}_{\mu \alpha \beta }=\left\{ _{\alpha \mu \beta }\right\} $. In
other words, in GR the inertial effect cannot be separated from the gravity, which is the origin of the covariance difficulty of the gravitational
energy-momentum.

\bigskip The motion of a classical particle follows the geodesic equation
\begin{equation}
\frac{du^{\mu }}{ds}+\left\{ {}_{\lambda \tau }^{\mu }\right\} u^{\lambda
}u^{\tau }=0.
\end{equation}%
\textit{\ }In the second term, $\left\{ {}_{\lambda \tau }^{\mu }\right\} $ represents a force, rather than a potential. The gravitational potential is
represented by the metric $g_{\mu \nu }$, as indicated by $g_{00}\sim
-\left( 1+2\phi \right) $ for the weak field approximation. As it is known, in electromagnetism and Yang-Mills theory, the connection represents the
gauge potential, while the curvature represents the force (field strength). The position error (wrong placement) of the potential and the field strength
is the origin of all the trouble and confusion in GR. For this reason, GR can never been formulated as a Yang-Mills theory.

The conclusion is that in GR the gravitational force is represented by the Christoffel connection rather than the curvature. It is a misunderstanding
that the gravitation is equivalent to a curved spacetime. A key point is that the Einstein equation (62) includes no dynamics of the curvature. It
indicates that the curvature of the spacetime is equivalent to a energy-momentum not a force. The energy-momentum of the matter field is the
source of the Christoffel connection rather than the source of the curvature. In both the active and passive aspects, the\ gravitational force
is the Christoffel connection rather than the curvature. The curvature tensor does not describe acceleration in a gravitational field but the
gradient of the acceleration (e.g. geodesic deviation). It represents only a tidal force, a effect of the gravity not itself. The mantra about
gravitation as curvature is a misnomer.

\section{Teleparallel gravity (TG)}

Teleparallel theory was considered by Einstein in 1928 as a possible geometrical set up for the unification of the electromagnetic and
gravitational fields [22].

Choose $\left\{ h^{a}{}_{\mu }\right\} $ to be a Lorentz basis and let $
R^{a}{}_{b\mu \nu }=0$. Using (53) and (45) we obtain \begin{equation}
\nabla _{\alpha }g_{\mu \nu }=-A{}_{\mu \nu \alpha }-A_{\nu \mu \alpha
}+h^{a}{}_{\mu }h^{b}{}_{\nu }\nabla _{\alpha }\eta _{ab},\text{ \ }
\end{equation}%
where $A{}_{\mu \nu \alpha }=h_{a\mu }h^{b}{}_{\nu }A^{a}{}_{b\alpha }$. For a Lorentz basis $h^{a}{}_{\mu }$, the corresponding connection (Lorentz
connection) $A^{a}{}_{b\mu }$ satisfies $A_{\nu \mu \alpha }=-A{}_{\mu \nu
\alpha }$, then we have
\begin{equation}
\nabla _{\alpha }g_{\mu \nu }=0.
\end{equation}
Using (43) we get
\begin{equation}
T_{b}{}^{a}{}_{c}=-f_{b}{}^{a}{}_{c}+A_{bc}{}^{a}-A_{b}{}^{a}{}_{c},\text{ \
\ }
\end{equation}
and then%
\begin{eqnarray}
&&T_{b}{}^{a}{}_{c}+T_{c}{}^{a}{}_{b}-T^{a}{}_{bc}  \notag \\
&=&f^{a}{}_{bc}-f_{b}{}^{a}{}_{c}-f_{c}{}^{a}{}_{b}+A_{bc}{}^{a}+A_{cb}{}^{a}+A^{a}{}_{bc}-A_{b}{}^{a}{}_{c}-A_{c}{}^{a}{}_{b}-A^{a}{}_{cb},%
\text{ \ }
\end{eqnarray}
where $A^{a}{}_{bc}=h_{c}{}^{\mu }A^{a}{}_{b\mu }$.

For the Lorentz connection $A^{a}{}_{b\mu }$, we have
\begin{equation*}
T_{b}{}^{a}{}_{c}+T_{c}{}^{a}{}_{b}-T^{a}{}_{bc}=f^{a}{}_{bc}-f_{b}{}^{a}{}_{c}-f_{c}{}^{a}{}_{b}+2A^{a}{}_{bc},
\end{equation*}
and then%
\begin{eqnarray}
A^{a}{}_{bc} &=&\frac{1}{2}\left(
-f^{a}{}_{bc}+f_{b}{}^{a}{}_{c}+f_{c}{}^{a}{}_{b}\right) +K_{b}{}^{a}{}_{c}
\notag \\
&=&A_{\left( 0\right) }^{a}{}_{bc}+K_{b}{}^{a}{}_{c},\text{ \ \ \ \ \ \ \ \
\ \ \ }
\end{eqnarray}
where
\begin{equation}
A_{\left( 0\right) }^{a}{}_{bc}=\frac{1}{2}\left(
-f^{a}{}_{bc}+f_{b}{}^{a}{}_{c}+f_{c}{}^{a}{}_{b}\right) .\text{ \ \ }
\end{equation}

The equation of a free particle as seen from a\emph{\ general Lorentz frame}
\begin{equation}
\frac{du^{a}}{ds}+A^{a}{}_{b\mu }u^{b}u^{\mu }=0\;\;\;\;
\end{equation}%
\emph{\ }takes the form [7]
\begin{equation}
\frac{du^{a}}{ds}+A_{\left( 0\right) }^{a}{}_{b\mu }u^{b}u^{\mu
}=K^{a}{}_{b\mu }u^{b}u^{\mu }.\text{ \ \ \ }
\end{equation}%
It is a force equation, with contortion $K^{a}{}_{b\mu }$\ playing the role of gravitational force. Contortion in this case acts as a force, which
similarly to the Lorentz force equation of electromagnetism, appears as an effective force term on the right-hand side of the motion equation of a
free-falling particle. The term $A_{\left( 0\right) }^{a}{}_{b\mu
}u^{b}u^{\mu }$ is non-covariant by its very nature and represents the inertial force coming from the frame non--inertiality, like centrifugal and
Coriolis force, is a fictitious force. In other words, whereas the gravitational effects are described by a covariant force $K^{a}{}_{b\mu }$,
the non--inertial effects of the frame are represented by an inertia--related Lorentz connection $A_{\left( 0\right) }^{a}{}_{b\mu }$. In
other words, the gravitational force acting on a particle or on a frame is due to the torsion tensor only. Notice that in (73), both inertial and
gravitational effects are included in the connection term $A^{a}{}_{b\mu
}u^{b}u^{\mu }$, like the case (66) in GR.

Since $A^{a}{}_{b\mu }$ is not Lorentz covariant, it is\ possible to find a local Lorentz frame in which $A^{a}{}_{b\mu }=0$\ and \ then $A_{\left(
0\right) }^{a}{}_{b\mu }=-K^{a}{}_{b\mu }$. This is just the equivalent principle: the inertial force $A_{\left( 0\right) }^{a}{}_{b\mu }$ cancels
the gravitational force $K^{a}{}_{b\mu }$.

Analogously to the Yang-Mills theories, we choose the quadratic function of the torsion (the external curvature)%
\begin{equation}
\mathcal{L=}\frac{1}{2}hS^{\rho \mu \nu }T_{\rho \mu \nu },\text{ \ \ \ \ \
\ \ }
\end{equation}%
as the gravitational Lagrangian, where $h=\det \left( h^{a}{}_{\mu }\right) $ and
\begin{equation}
S^{\rho \mu \nu }=\frac{1}{2}\left( K^{\mu \nu \rho }-g^{\rho \nu }T^{\sigma
\mu }{}_{\sigma }+g^{\rho \mu }T^{\sigma \nu }{}_{\sigma }\right) .\text{ \ }
\end{equation}%
The variational principle gives the field equation of $h^{a}{}_{\mu }$:
\begin{equation}
\partial _{\sigma }\left( hS_{\lambda }{}^{\sigma \rho }\right) =\frac{1}{2}%
h\left( t_{\lambda }{}^{\rho }+\Theta _{\lambda }{}^{\rho }\right) ,\text{ \
\ \ \ \ \ \ \ \ \ \ \ \ \ \ \ \ \ }
\end{equation}
where $ht_{\lambda }{}^{\rho }$ is the energy-momentum tensor of gravitation plus inertial effects and $h\Theta _{\lambda }{}^{\rho }$ is the source
energy-momentum tensor [7, 23]. (77) is the same as Einstein's equation. Considering that the teleparallel force equation (73) and the geodesic
equation (66) of GR are formally the \textbf{same}, the teleparallel description of the gravitational interaction is equivalent to the
description of GR.

There exists a solution $A^{a}{}_{b\mu }=0$ for the equation $R^{a}{}_{b\mu \nu }=0$. Since the curvature is the field strength of the connection, its
vanishing implies that the connection $A^{a}{}_{b\mu }$ must be purely inertial [15] and\ has nothing to do with gravitation.

In this case the tetrad $h^{b{}}{}_{\mu }$ depends only the teleparallel potential $B^{a{}}{}_{\mu }$:

\begin{equation}
h^{a{}}{}_{\mu }=h^{\left( 0\right) a{}}{}_{\mu }=e^{a{}}{}_{\mu
}+B^{a{}}{}_{\mu },\text{ \ }
\end{equation}%
which includes only the pure gravity $B^{a{}}{}_{\mu }$ without the inertial force $A^{a}{}_{b\mu }$. The connection $\Gamma ^{\rho }{}_{\mu \nu }$ and
the tetrad $h^{a}{}_{\mu }$ are not independent each other any more, but related:

\begin{equation}
\Gamma ^{\rho }{}_{\mu \nu }=h_{a}^{\left( 0\right) }{}^{\rho }\partial
_{\mu }h^{\left( 0\right) a}{}_{\nu }.\text{ \ \ \ \ \ \ \ }
\end{equation}
The torsion $T^{a}{}_{\mu \nu }$ depends only on the tetrad $h^{a}{}_{\mu
}=h^{\left( 0\right) a{}}{}_{\mu }$ and is equal to the field strength $ F^{a}{}_{\mu \nu }$ of the translation potential $B^{a}{}_{\mu }$:
\begin{equation}
T^{a}{}_{\mu \nu }=\partial _{\mu }h^{\left( 0\right) a}{}_{\nu }-\partial
_{\nu }h^{\left( 0\right) a}{}_{\mu }=\partial _{\mu }B^{a}{}_{\nu
}-\partial _{\nu }B^{a}{}_{\mu }=F^{a}{}_{\mu \nu }.\text{ \ \ }
\end{equation}
The equation (48) gives
\begin{equation}
f^{c}{}_{ab}=h_{a}^{\left( 0\right) }{}^{\mu }h_{b}^{\left( 0\right)
}{}^{\nu }(\partial _{\nu }h^{\left( 0\right) c}{}_{\mu }-\partial _{\mu
}h^{\left( 0\right) c}{}_{\nu })=h_{a}^{\left( 0\right) }{}^{\mu
}h_{b}^{\left( 0\right) }{}^{\nu }T^{c}{}_{\nu \mu }.\text{ \ \ }
\end{equation}
and%
\begin{equation}
f^{c}{}_{ab}=h_{a}^{\left( 0\right) }{}^{\mu }h_{b}^{\left( 0\right)
}{}^{\nu }(\partial _{\nu }B^{c}{}_{\mu }-\partial _{\mu }B^{c}{}_{\nu
})=h_{a}^{\left( 0\right) }{}^{\mu }h_{b}^{\left( 0\right) }{}^{\nu
}F^{c}{}_{\nu \mu }.\text{ \ \ \ \ \ \ \ }
\end{equation}
This is a\ pure gravity with out inertial.

The equation (43) indicates that the general tetrad $h^{b{}}{}_{\mu }$ is given by a combination of the connections $A^{b}{}_{c\mu }$ and $
B^{b{}}{}_{\mu }$. In other words, $A^{b}{}_{c\mu }$ is not independent of $h^{b{}}{}_{\mu }$ and there is no equation of it. In the gauge $
A^{a}{}_{b\mu }=0$, the field equation (77) is just the motion equation of gravitational field $B^{b{}}{}_{\mu }$. In this case the frame is called
proper. So the gauge $A^{a}{}_{b\mu }=0$ corresponds to choosing an accelerated reference system to cancel the inertial force in Galileo-Newton
mechanics. And then $A_{\left( 0\right) }^{a}{}_{b\mu }$ correspond the inertial force.

In special relativity, the anholonomy of the frames is entirely related to the inertial forces $A^{a}{}_{b\mu }$ present in those frames. The preferred
class of frames, the inertial class is characterized by the absence of inertial forces $A^{a}{}_{b\mu }$, and consequently by holonomic frames.
Similarly, in the presence of gravitation $B^{a}{}_{\mu }$ there is also a preferred class of frames i.e.\ proper frames whose anholonomy is related to
gravitation $B^{a}{}_{\mu }$\ only, not with inertial effects $A^{a}{}_{b\mu }$. This class of frames, therefore, reduces to the inertial class when
gravitation is switched off. [24]

In teleparallel gravity the Lorentz connection $A^{a}{}_{b\mu }$ keeps its special-relativistic role of representing inertial effects only and is
neither a true field variable nor a genuine gravitational connection. The gravitational field is fully represented by the translation connection $%
B^{a}{}_{\mu }$, the non-trivial part of the tetrad $h^{\left( 0\right)
a{}}{}_{\mu }$. Its\ field strength $F^{a}{}_{\mu \nu }=\partial _{\mu
}B^{a}{}_{\nu }-\partial _{\nu }B^{a}{}_{\mu }$\ is just the anholonomy $
f^{c}{}_{ab}=h_{a}^{\left( 0\right) }{}^{\mu }h_{b}^{\left( 0\right)
}{}^{\nu }F^{c}{}_{\nu \mu }$\ of the tetrad $h^{\left( 0\right) a{}}{}_{\mu
}$. [25]. The translational connection $B^{a}{}_{\mu }$ represents gravitation only, without inertial effects. Furthermore, it is also a
genuine gravitational connection. Its connection behavior under gauge translations is related uniquely to its gravitational content.

It should be emphasized that the Lagrangian (75) and the field equation (77) are covariant under coordinate transformations, local translations as well
as local Lorentz transformations. However, whenever the condition $ A^{a}{}_{b\mu }=0$ is chosen, we attain a fixed Lorentz gauge and then
cannot talk about the issue of Lorentz covariance. This case is similar to the fixed gauge, for example, the Coulomb gauge in electrodynamics.
Therefore, the case $A^{a}{}_{b\mu }=0$ is only a fixed gauge in TG not the theory itself. In literature, however, it is often called " teleparallel
equivalent of general relativity" (TEGR) or \textquotedblleft pure tetrad teleparallel gravity" and then gives rise to debate of covariance property
of the theory [26-31].

However, some recent works [{\small 27-29}] still try  to exhibit local Lorentz symmetry of the TEGR. It is shown that teleparallel gravity and its
generalizations can be formulated as fully invariant (both coordinate and local Lorentz) theories of gravity. At the same time, several
inconsistencies of TG with local Lorentz symmetry are proposed. A analysis of the problems regarding a surface term is presented [26].

In spite of the equivalence, there are conceptual differences between GR and TG. GR cannot separate gravitation from inertial effects, but\ that becomes
possible by acknowledging the frame $h^{a}{}{}_{\mu }$ in TG [21]. The association of the pure-gauge Lorentz connection with inertial effects has
been clarified in depth in the context of TG [13]. The most significant is the different character of the fundamental field of the theories. Whereas it
is a metric tensor $g_{\mu \nu }$ in GR, it is a translational gauge potential $B^{a}{}_{\mu }$ in TG.

In GR, geometry replaces the concept of force, and the trajectories are determined by geodesics. TG, on the other hand, attributes gravitation to
torsion (anholonomy). Torsion, however, accounts for gravitation not by geometrizing the interaction, but by acting as a force. In consequence,
there are no geodesics in teleparallel gravity, only force equations analogous to the Lorentz force equation of electrodynamics. This is actually
an expected result because, like electrodynamics, teleparallel gravity is a gauge theory. [33]

In a new formulation of TEGR the Cartan Geometry and the Cartan connection are used as an alternative approach to describe TEGR as a gauge theory of
translations. [34-37].

In TG\emph{\ torsion }appears as an alternative to curvature in the description of the gravitational field, and is consequently related with the
same degrees of freedom of gravity. This interpretation is completely different from that appearing in more general theories, like
Einstein--Cartan and gauge theories for the Poincare and the affine groups. In these theories, curvature and torsion are considered as independent
fields, related with different degrees of freedom of gravity, and consequently with different physical phenomena.

\section{Symmetric teleparallel gravity (STG)}

The standard tetrad formalism of GR has been teleparallelised in the metric-affine gauge theory [38,39], but it still is possible to realise in a
more trivial geometry. In the case $R^{\alpha }{}_{\beta \mu \nu
},=0,T{}_{\beta \alpha \mu }=0$, \ GR is\ formulated in a flat and torsion-free spacetime and obtains a simplifying geometrical frameworks. We can
introduce a class of its generalisations in the\ trivially connected geometry, i.e. with $\Gamma ^{\alpha }{}_{\mu \nu }=0$. We directly deal
with the spacetime metric $g_{\mu \nu }$ and avoid introducing a frame field. This is STG model, in which the metric $g_{\mu \nu }$ represents the
gravitational potential and its derivatives $\left\{ _{\gamma \mu }^{\alpha
}\right\} $ represents the gravitational field strength. In other words, we come back to the original approach of Einstein but in a\emph{\ }trivially
connected geometry abandoning the curvature. This representation of GR covariantizes (and hence legitimizes) the usual coordinate calculations. The
gravitational interaction effects, via the nonmetricity $Q_{\mu \alpha \beta
}$, have a character much like a Newtonian force with a potential $g_{\mu\nu }$.

The choice\ $\Gamma {}_{\mu \alpha \beta }=0$ leads to
\begin{equation}
Q_{\mu \alpha \beta }=\partial _{\mu }g_{\alpha \beta },\text{ \ \ \ }
\end{equation}
and
\begin{equation}
L_{\mu \alpha \beta }=-\left\{ _{\alpha \mu \beta }\right\} =-\frac{1}{2}
\left( \partial _{\alpha }g_{\mu \beta }+\partial _{\beta }g_{\alpha \mu
}-\partial _{\mu }g_{\alpha \beta }\right) ,\text{ \ }
\end{equation}

The Christoffel (or Levi-Civita) connection is canceled by the disformation. However, the curvature of the Christoffel connection or the curvature of the
metric does not vanish. It is not the whole curvature, but only a part of the later, since the Christoffel connection is only a part of the whole
connection as shown by (56). Gravity is described by neither the torsion nor the curvature of the affine geometry. Based on this property, it proposes a
simpler geometrical formulation of GR which is oblivious to the affine structure and fundamentally depriving gravity of any inertial character.
This theory has nothing to do with Lorentz groups and inertial effects and therefore is a purified gravity theory [18].\emph{\ }It can be described by
the Einstein ($\Gamma \Gamma $) Lagrangian
\begin{equation}
\mathcal{L}=\sqrt{-g}g^{\mu \nu }\left( \left\{ _{\gamma \mu }^{\alpha
}\right\} \left\{ _{\nu \alpha }^{\gamma }\right\} -\left\{ _{\gamma \alpha
}^{\alpha }\right\} \left\{ _{\mu \nu }^{\gamma }\right\} \right) ,
\end{equation}%
which is nothing but the quadratic piece of the Hilbert Lagrangian. It differs from the Ricci scalar only by a total derivative so that, written in
terms of the metric, both Lagrangians are equivalent. We want to emphasize that, in terms of the STG geometry this Lagrangian is viewed as covariant
[17]. It is also the unique quadratic form covariant with respect to translations. Thus, we arrive at an improved version of GR where the
boundary term is absent, the connection can be fully trivialised and, thus, the inertial effect of the connection has completely disappeared.

The Lagrangian (85) yields the explicit form of the Einstein field equation
\begin{equation}
\partial _{\alpha }\left( \sqrt{-g}\Pi _{\lambda }^{\alpha \mu \nu \rho
\sigma }\left( g\right) \left\{ _{\rho \sigma }^{\lambda }\right\} \right) =
\sqrt{-g}\left( t^{\mu \nu }+T^{\mu \nu }\right) ,
\end{equation}
where
\begin{eqnarray}
\Pi _{\lambda }^{\alpha \mu \nu \rho \sigma }\left( g\right)  &=&\frac{1}{2}
\{\left( g^{\nu \rho }g^{\sigma \mu }+g^{\sigma \nu }g^{\mu \rho }-g^{\rho
\sigma }g^{\nu \mu }\right) \delta _{\lambda }^{\alpha }+\left( g^{\mu \nu
}g^{\rho \alpha }-2g^{\alpha \nu }g^{\rho \mu }\right) \delta _{\lambda
}^{\sigma }  \notag \\
&&+\left( g^{\rho \sigma }g^{\nu \alpha }-g^{\nu \rho }g^{\sigma \alpha
}\right) \delta _{\lambda }^{\mu }+\left( g^{\sigma \alpha }g^{\rho \mu
}-g^{\rho \sigma }g^{\alpha \mu }\right) \delta _{\lambda }^{\nu }\},\text{
\ \ \ \ }
\end{eqnarray}
\begin{eqnarray}
t^{\mu \nu } &=&\left( \frac{1}{2}g^{\mu \nu }g^{\tau \sigma }-g^{\mu \tau
}g^{\nu \sigma }\right) \left( \left\{ _{\lambda \tau }^{\rho }\right\}
\left\{ _{\sigma \rho }^{\lambda }\right\} -\left\{ _{\rho \gamma }^{\gamma
}\right\} \left\{ _{\tau \sigma }^{\rho }\right\} \right)   \notag \\
&&-g^{\tau \rho }g^{\mu \sigma }\left( \left\{ _{\sigma \lambda }^{\lambda
}\right\} \left\{ _{\tau \rho }^{\nu }\right\} -\left\{ {}_{\lambda \tau
}^{\nu }\right\} \left\{ _{\rho \sigma }^{\lambda }\right\} +\left\{
_{\lambda \sigma }^{\nu }\right\} \left\{ _{\tau \rho }^{\lambda }\right\}
-\left\{ _{\sigma \tau }^{\lambda }\right\} \left\{ _{\rho \lambda }^{\nu
}\right\} \right) ,\text{ \ \ \ }
\end{eqnarray}
\begin{equation}
T^{\mu \nu }=-\frac{2}{\sqrt{-g}}\frac{\delta \mathcal{L}_{\text{m}}}{\delta
g_{\mu \nu }}.\text{ \ \ \ \ \ \ \ \ \ \ \ \ \ \ \ \ \ \ \ \ \ \ \ \ \ \ \ \
\ \ \ \ \ \ \ \ \ \ \ \ \ }
\end{equation}

The assumption $\Gamma ^{\alpha }{}_{\mu \nu }=0$ means $T^{c}{}_{\mu \beta }=0$ and $R^{c}{}_{a\mu \beta }=0$. Then (50) becomes
\begin{equation}
\partial _{\nu }B^{a}{}_{\mu }-\partial _{\mu }B^{a}{}_{\nu }=A^{a}{}_{b\mu }B^{b}{}_{\nu }-A^{a}{}_{b\nu }B^{b}{}_{\mu }.\text{ \ \ \ }
\end{equation}%
One can find  that although $T^{a}{}_{b\mu \nu }=0,R^{a}{}_{b\mu \nu }=0,$
the gravitational field $F^{a}{}_{\nu \mu }=\partial _{\nu }B^{a}{}_{\mu }-\partial _{\mu }B^{a}{}_{\nu }\neq 0.$ However, since the connection $
A^{a}{}_{b\mu }$ is not a tensor, it can be transformed such that makes $F^{a}{}_{\nu \mu }=0.$ This is just the so called equivalence principle.

There exists a solution $A^{a}{}_{b\mu }=0$ for the equation $R^{a}{}_{b\mu \nu }=0$. In this case the tetrad $h^{a{}}{}_{\mu }=h^{\left( 0\right)
a{}}{}_{\mu }=e^{a{}}{}_{\mu }+B^{a{}}{}_{\mu }$ and then the metric $g_{\mu\nu }=\eta _{ab}h^{a{}}{}_{\mu }h^{b{}}{}_{\nu }$ depend only the
teleparallel potential $B^{a{}}{}_{\mu }$. In other words, they includes only the pure gravity $B^{a{}}{}_{\mu }$ without the inertial force $
A^{a}{}_{b\mu }$. Consequently, the metric $g_{\mu \nu }=\eta
_{ab}h^{a{}}{}_{\mu }h^{b{}}{}_{\nu }$ and the disformation $L_{\mu \alpha
\beta }$ depend only the pure gravity $B^{a{}}{}_{\mu }$.

\ In STG that the connection can be exactly cancelled by a diffeomorphism. The gauge in which the connection is trivialised is called the coincident
gauge [18]. The origins of the tangent space and the spacetime coincide. Reducing the general linear symmetry to the diffeomorphic symmetry of the
coordinate transformations suggests a new foundation for the gravitational geometry.

\section{The energy-momentum of gravitational fields}

The definition of an energy-momentum density for the gravitational field is one of the oldest and most controversial problems of gravitation and has a
long and complicated history. In order to solve this problem, numerous formalisms were developed to derive the conservation laws. There are two
classes of conservation laws in gravity theories. One class of conservation laws is formulated solely in terms of the dynamic variables that describe
the gravitational field itself. These variables characterize the geometry of spacetime without involving additional structures of physical
(non-geometrical) nature. A unified framework has been developed to discuss Noether identities and conservation laws for a wide range of gravitational
theories [40].

Another class of conservation laws lead to conserved charges and turn out to be true scalars. Deriving such conservation laws needs to deal with, besides
the gravitational field variables, additional physical structures such as vector fields. A vector field can be associated with a current that is
conserved under some conditions. In GR this vector is a Killing vector, the current is conserved. This fact establishes a remarkable relation between
the symmetries of the spacetime and the conserved currents generated by these symmetries. Physically, this vector field is usually related to the
reference frame motion of an observer.

All fundamental fields have a well-defined local energy-momentum density. It is then expected that the same should happen to the gravitational field. As
a true field, gravity should have its own local energy-momentum density. However, it is usually asserted that such a density can not be locally
defined because of the equivalence principle [1]. In the context of general relativity, no tonsorial expression for the gravitational energy-momentum
density can exist. Consequently, many attempts to identify an energy-momentum density for the gravitational field lead to complexes that
are not true tensors. The first of such attempt is Einstein's expression for the energy-momentum density of the gravitational field which is nothing but
the canonical expression obtained from Noether's theorem [16, 41]. According to this theorem, if the action of a physical system consisting of fields $
\psi $ is invariant under a transformation of the fields $\psi $, the corresponding Lagrangian $\mathcal{L}$ determines a conserved current. For
the translation group the conserved current is the energy-momentum
\begin{equation}
T^{\mu }{}_{\nu }=\frac{\partial \mathcal{L}}{\partial \left( \partial _{\mu
}\psi \right) }\partial _{\nu }\psi -\delta _{\nu }^{\mu }\mathcal{L}.\text{
\ \ \ \ \ \ \ \ \ }
\end{equation}

In the Yang-Mills theory the conserved current is the isospin, the weak, or the colour current. The non-covariant property of non-abelian gauge fields
is a pervasive problem in gauge field theories. The conserved current of a gauge field $A^{i}{}_{\mu }$\ is
\begin{equation}
\ J_{\left( A\right) i}{}^{\mu }=-\frac{\partial \mathcal{L}\left( A\right)
}{\partial \left( \partial _{\mu }A^{j}{}_{\nu }\right) }%
c_{i}{}^{j}{}_{k}A^{k}{}_{\nu },\text{ \ \ \ \ \ \ }
\end{equation}%
where $c_{i}{}^{j}{}_{k}$ is the structure constant of the transformation group $G$. Due to the connection nature of $A^{k}{}_{\nu }$, the current $
J_{\left( A\right) i}{}^{\mu }$ is not covariant under the gauge transformation.

In GR the Einstein expression for the energy-momentum density of the gravitational field
\begin{equation}
t^{\mu }{}_{\nu }=\frac{\partial \mathcal{L}_{\left( g\right) }}{\partial
\left( \partial _{\mu }g_{\rho \sigma }\right) }\partial _{\nu }g_{\rho
\sigma }-\delta _{\nu }^{\mu }\mathcal{L}_{\left( g\right) },\text{ \ \ }
\end{equation}
is given by (88). Due to the ordinary derivative $\partial _{\nu }g_{\rho \sigma }$, $t^{\mu }{}_{\nu }$ is not covariant under the coordinate
transformations.

Einstein vigorously defended [42] the use of the pseudotensor (93), which was also adopted, amongst many others, by Dirac in his textbook [43].

In TG, if $B^{a}{}_{\mu }$ is chosen as the field variable, the energy-momentum density of the gravitational field is
\begin{eqnarray}
j_{\alpha }{}^{\beta } &=&\frac{\partial \mathcal{L}_{\left( B\right) }}{\partial \left( \partial _{\beta }B^{b{}}{}_{\tau }\right) }\partial
_{\alpha }B^{b{}}{}_{\tau }-\mathcal{L}_{\left( B\right) }\delta _{\alpha
}{}^{\beta }  \notag \\
&=&2\frac{\partial \mathcal{L}_{\left( B\right) }}{\partial \left( \partial
_{\beta }g_{\mu \nu }\right) }h{}_{b\nu }\partial _{\alpha }B^{b{}}{}_{\mu }-
\mathcal{L}_{\left( B\right) }\delta _{\alpha }{}^{\beta },\text{ \ \ \ }
\end{eqnarray}%
Due to the connection nature of $B^{b{}}{}_{\mu }$ the energy-momentum density $j_{\alpha }{}^{\beta }$ is not covariant under the gauge
translations but is covariant under general spacetime coordinate transformation. However, if $h^{a}{}_{\mu }$ is chosen as the field
variable, the energy-momentum density of the gravitational field is
\begin{eqnarray}
ht_{\sigma }{}^{\rho } &=&\frac{\partial \mathcal{L}_{\left( h\right) }}{%
\partial \left( \partial _{\rho }h^{a{}}{}_{\mu }\right) }\partial _{\sigma
}h^{a{}}{}_{\mu }-\mathcal{L}_{\left( h\right) }\delta _{\sigma }{}^{\rho }
\notag \\
&=&\frac{1}{2}hh_{a}{}^{\mu }\partial _{\sigma }h^{a{}}{}_{\nu }S_{\mu
}{}^{\nu \rho }-\mathcal{L}_{\left( h\right) }\delta _{\sigma }{}^{\rho },
\text{ \ \ \ }
\end{eqnarray}%
which is covariant under the gauge translations but not covariant under general spacetime coordinate transformation. At the same time, both $%
j_{\alpha }{}^{\beta }$ and $t_{\sigma }{}^{\rho }$ are covariant under global Lorentz transformations but not covariant under local Lorentz
transformations.

In STG if $g_{\mu \nu }$ is chosen as the field variable
\begin{equation}
T^{\mu }{}_{\nu }=\frac{\partial \mathcal{L}_{\left( g\right) }}{\partial
\left( \partial _{\mu }g_{\rho \sigma }\right) }\partial _{\nu }g_{\rho
\sigma }-\delta _{\nu }^{\mu }\mathcal{L}_{\left( g\right) },\text{ \ \ }
\end{equation}%
which is covariant under local translations, Local Lorentz transformations and general coordinate transformations.

STG covariantizes (and hence legitimizes) the usual coordinate calculations. The associated energy-momentum density (96) is just the Einstein
pseudotensor (93), but in this novel geometric representation it is a true tensor. The result, the Einstein \emph{pseudotensor}, from the STG viewpoint
is a covariant object. The coordinate dependence of the energy-momentum density is now elevated from a choice of reference frame to a
\textquotedblright gauge\textquotedblright\ choice of geometry\emph{. }Thus the associated energy-momentum density is a covariant (but geometry gauge
dependent) tensor.

In the gauge context, the existence of a tensor expression for the gravitational energy-momentum density turns out to be possible. The absence
of such expression is attributed to the GR description of gravitation, which is not the appropriate framework to deal with this problem. The basic reason
is that both gravitational and inertial effects are mixed in the connection, and cannot be separated. Then the energy-momentum density includes both the
part of gravity and the part of the inertial effects. Since the inertial effects are essentially non-tensorial, the quantity defining the
energy-momentum density of the gravitational field in GR always shows up as a non-tensorial object.

On the other hand, owing to the possibility of separating gravitation from inertial effects in TG, it is possible to write down an energy-momentum
density for gravitation only, excluding the contribution from inertia. Such quantity is a tonsorial object. It is used to compute the energy of any
gravitational system and always gives the physical result, no matter the coordinates or frames used to make the computation [44].

In TG the energy-momentum currents are covariant under general coordinate transformations and global Lorentz transformations of the frame (tetrad
field). However, they are not covariant under local Lorentz transformations. As a consequence, the corresponding energy and momentum densities and also
the total conserved quantities in a given spacelike region are\ coordinate independent. However, they do depend on the chosen frame [45,46].

Another class of conservation laws lead to conserved charges. Deriving such conservation laws needs to deal with additional physical structures such as
vector fields. It is known that conserved currents can be associated with spacetime diffeomorphisms represented by the Killing vector fields of the
symmetries. A systematic derivation of a general expression for conserved currents has been presented [47]. Such currents are associated with a
arbitrarily given vector field on the spacetime manifold. They are invariant under both coordinate and local Lorentz transformations. The conservation
law holds \textquotedblleft on-shell\textquotedblright , i.e. on every solution of the coupled system of the gravitational and matter field
equations. An applications of the result is the computation of the total mass and angular momentum for the solutions in the gravitational theories
[48].

The interpretation of the vector field associated with conserved currents is an important geometrical and physical issue. For example, when it is
timelike, the corresponding charge has the meaning of the energy of the gravitating system with respect to an observer moving along the integral
lines of the vector. In this way, the dependence of the conserved charges on the vector describes the usual dependence of the energy of a system on the
choice and on the dynamics of a physical observer.

In order to get\ physically meaningful conserved charges, it is crucial to choose an appropriate (or preferred) frame, in which the inertial effects
are absent. The coordinate counterpart of this property is well known. Similarly to the choice of the preferred frame, the choice of this
coordinate system is crucial in the sense that it does not introduce spurious effects in the calculation of the energy.

\section{Modified gravity theories}

In the last two decades, cosmological observations have triggered investigations seeking for theories beyond GR, mainly motivated by the three
fundamental missing ingredients of the standard cosmological model: dark matter, dark energy and inflation. Nowadays, extended or alternative
formulations of general relativity are investigated with the purpose of solving cosmological problems, establishing a possible quantum theory of
gravity, and addressing unsolved issues of the standard formulation of general relativity. A plethora of modified or extended theories of gravity
has been proposed and studied [49].

Such generalized theories are interesting in modelling of dark energy, of which the \textbf{origin} is theoretically unknown but the properties can be
experimentally tested. A large number of gravitational dark energy models have been proposed previously, recall e.g. [50-53].

Instead of changing the source side of the Einstein field equations, a geometrical approach\ can be taken into account to fit the missing
matter-energy of the observed universe. The equivalent descriptions of GR with curvature, torsion and non-metricity represent different alternative
starting points to modified gravity theories once the corresponding Lagrangians are promoted to arbitrary functions $f(R)$, $f(T)$, and $f(Q)$.

The actions of GR, TG and STG differ by a non-trivial boundary term. The boundary term does not contribute to the field equations. Therefore, GR, TG
and STG are equivalent. However, $f(R)$, $f(T)$ and $f(Q)$ theories are not equivalent. This is due to that an arbitrary function of a boundary term is,
in general, no longer a boundary term.

Modified gravity theories based on $f(R)$ [54-56] and $f(T)$ [50,57] have been widely studied in the literature. We therefor concentrate on a brief
overview of $f(Q)$ theories.

Until two decades ago, the $f(Q)$ theory is less studied. However it has gained more and more attention recently. One of the essential features of
the $f(Q)$ theory is that, unlike $f(R)$ theory, it can separate gravity from the inertial effects. It is also worth mentioning that in contrast to $
f(R)$ and $f(T)$ gravity, $f(Q)$ gravity is free from pathologies. The field equations in $f(R)$ gravity are fourth- order, but they are of second- order
in $f(Q)$ gravity. The $f(T)$-gravity models suffer from strong coupling problems when considering perturbations around a FLRW background [58], these
are absent in the case of $f(Q)$-gravity [59].

Thus, $f(Q)$ gravity forms a novel starting point for various modified gravity theories. There is a number of recent studies of modified STG models
[18,59,60] and their applications in cosmology [61]. A cosmology of STG has been developed [60]. In this theory the accelerating expansion is an
intrinsic property of the universe geometry without need of either exotic dark energy or extra fields. The role of dark energy is played by the
geometry itself and then is endowed with intrinsic character of the spacetime. Now we are returning to the original idea of Einstein and
Wheeler: gravity is a geometry [1,62,].

At the cosmological background level, models based on $f(Q)$ are indistinguishable from $f(T)$ models, but their perturbations have
distinctive properties from the ones of $f(T)$ theories. At the small-scale quasi-static limit the predictions of $f(Q)$ and $f(T)$ models coincide, but
at larger scales $f(Q)$ models generically propagate two scalar degrees of freedom that are absent in the case of $f(T)$.

Various work indicate that the $f(Q)$ theory is one of the promising alternative formulations of gravity. Without the restriction to quadratic
terms, cosmological solutions based on Noether symmetries were derived for generic first derivative order non-metric actions [63]. The energy
conditions restricting the parameters of two $f(Q)$ models of the universe has been analyzed [64]. A class of $f(Q)$ theories with the non-minimally
coupled $Q$ to the matter sector has been constructed [65,66]. In scalar-nonmetricity theories of gravity [67-69], the resemblances of these
theories with scalar-curvature and scalar-torsion models and considered conformal transformations have been explored. The cosmological behavior of $
f(Q)$ theory at the background level has been investigated from a dynamic system perspective [70]. The integrability of the model has been examined by
employing the method of singularity analysis techniques [70].

Some theoretical predictions on linear cosmological observables from a specific class of $f(Q)$-gravity models have been presented and suggested to
be tested extensively against data. Observational constraints on the background behavior of several $f(Q)$ models have been performed by testing
against various current background data such as Type Ia Supernovae, Pantheon data, Hubble data, etc. [71, 72]. These studies conclude that viable $f(Q)$
models resemble the GR-based model viz. $\Lambda $CDM model. The confrontation with observation data reveals that the scenario is
statistically preferred comparing to $\Lambda $CDM cosmology at some cases [73].

The modified Gravitational Waves (GWs) propagation in STG has been investigated [74-78], the possible new parity-violating signatures have been
explored [79, 80]. The relevance of the modified Newtonian limit in $f(Q)$ gravity to dark matter phenomenology has been investigated in [81, 82].

A Bayesian statistical analysis using redshift space distortions data is performed to test a $f(Q)$ model. It reveals that the $\sigma _{8}$ tension
between Planck and Large Scale Structure data can be alleviated within this framework [83].

Cosmography has been used in $f(Q)$ gravity to investigate the cosmic expansion history of the Universe. The constraints on the Hubble constant $
H_{0}$ and the cosmographic functions were derived using the MCMC analysis [84].

The $f(Q)$ theories in the coincident gauge do not have the usual gauge invariance of cosmological perturbations. They are no longer
diffeomorphism-invariant, the equations do change under a diffeomorphism. However, maximally symmetric backgrounds retain a gauge symmetry given by a
restricted diffeomorphism.

It is found out that sometimes the coincident gauge conflicts with the selected coordinate system based on symmetry [85].To circumvent this
problem\ the $f(Q)$\ theory is formulated in a covariant way which consider both the metric and the affine connection as fundamental variables. A method
is proposed to search for suitable affine connections in $f(Q)$ theory for a given metric ansatz.

In a recent paper [86], starting from the original Einstein action, the $ \Gamma \Gamma $ action, a new formulation of modified theories of gravity is
proposed as has been done in [87]. This formulation is based on the metric alone and does not require more general geometries. It is equivalent to the $
f(Q)$ gravity\ at the level of the action and the field equations, provided that appropriate boundary terms are taken into account. It can also match up
with $f(R)$ gravity and be extended to $f(T)$ gravity. Under some conditions the field equations of these three theories are
indistinguishable. In this approach boundary terms play a important role. The main result is to identify three boundary terms with unusual properties
that allows us to construct one general family of modified gravity theories which will contain the various models as limiting cases.

\section{Discussion of some concepts}

\subsection{ Equivalence Principle}

Galileo's experiment on the Leaning Tower of Pizza indicates that the
free-fall of different objects is independent of their internal composition.
This result is elevated to the Equivalence Principle\ as one of the
foundations of GR, where Einstein's elevators displaces Galileo's falling
objects. In TG and STG this principle\ means that the gravitational force
can be canceled by a Lorentz connection in the same way as the gravitational
force being canceled by the inertial force in Newton mechanics.

The principle of equivalence was not new in Newton's theory of gravitation.
New was Einstein's extension to all of physics. Einstein wrote\ [88]: "...
we shall therefore assume the complete physical equivalence of a
gravitational field and a corresponding acceleration of the reference
system.\textquotedblright\ However, this notion makes mathematical sense
only for bodies having the same 4-velocity and from a physical point of view
accelerations are absolute.

Einstein argued that the equivalence of inertia and gravitation made
acceleration against absolute space obsolete. Only relative acceleration of
bodies was supposed to\ be observable. All inertial forces could then also
be interpreted as gravitational ones. However, these gravitational forces
have no sources and are generated by coordinate transformations and then can
not be equivalent to the real gravitational fields.

For the modern physicist, acceleration against Newton's absolute space is no
longer so implausible as it appeared a century ago. The vacuum of spacetime
is filled with a non-zero Higgs field and fluctuations of other fields.

As early as in 1960, John Synge confesses [89]: \textquotedblleft . . . I
have never been able to understand this Principle.\textquotedblright\ \
\textquotedblleft Does it mean that the effects of a gravitational field are
indistinguishable from the effects of an observer's acceleration? If so, it
is false. This is an absolute property, it has nothing to do with any
observer's worldline. ... The Principle of Equivalence performed the
essential office of midwife at the birth of general relativity, but, as
Einstein remarked, the infant would never have got beyond its long-clothes
had it not been for Minkowski's concept. I suggest that the midwife be now
buried with appropriate honours and the facts of absolute space-time
faced.\textquotedblright\

The theoretical relevance of the equivalence principle is mere the
indication of the geometric nature of gravity. The modern view was stated by
Bondi at the occasion of the Centenary of Einstein's birthday in 1979. He
wrote [90]:

\textquotedblleft From this point of view, Einstein's elevators have nothing
to do with gravitation; they simply analyze inertia in a perfectly Newtonian
way. Thus, the notion of general relativity does\ not in fact introduce any
postNewtonian physics;\ it simply deals with coordinate transformations.
Such a formalism may have some convenience, but physically it is wholly
irrelevant. It is rather late to change the name of Einstein's theory of
gravitation, but general relativity is a physically meaningless phrase that
can only be viewed as a historical memento of a curious philosophical
observation.\textquotedblright

Generalizations of GR have been introduced to account for modification of
GR, both at early and late phases of the Universe evolution. Cosmological
Inflation, Dark Matter and Dark Energy represent the main topics of this
strive. From the other side, GR is not a fundamental theory of physics
because it should require the inclusion of quantum effects. It is then
natural to ask whether the Equivalence Principle still holds in the
framework of any modified gravity approach aimed to enclose quantum physics
under the standard of gravitational interaction.

\subsection{General Covariance Principle}

Another foundations of GR is the General Covariance Principle. Due to the
intrinsic relation of gravitation with spacetime geometry, there is a deep
relationship between covariance and the equivalence. As the starting point
of GR, the principle of general covariance is considered as an \textbf{active%
} version of the equivalence principle. Making a special relativity equation
covariant, it is possible to obtain its form in the presence of gravitation.
This description of the general covariance principle refers to its usual
holonomic version. However, in this case the general covariance alone is
empty of any physical content. As early as 1917, Kretschmann criticized it
and asserted that any spacetime theory could be formulated in a generally
covariant way without any physical principle [91]. An alternative version of
the principle can be obtained in the context of anholonomic frames. Instead
of requiring the covariance under a general coordinate transformation, the
anholonomic version requires the covariance under local Lorentz
transformations of frames.

For some time, the idea of a reference frame was ambiguous because it was
mixed with the idea of a coordinate system. Nowadays, there is a\ clear
distinction between them. Coordinates are used to assign four numbers to
events on spacetime. In the mathematical language a coordinate system is a
chart on the spacetime manifold, which is meaningless physically. On the
other hand\ a reference system can be considered as a simplified geometric
representation of the quantized measuring instruments. It determine the
metric and various physical objects. In general, the base of a coordinate
system is holonomic, while a reference frame is anholonomic.

Covariance of the field quantities under local Lorentz transformations imply
that the measurement of these quantities is the same in inertial and
non-inertial frames. However, this is not an expected feature of concepts
such as energy, momentum and angular momentum, which cannot be covariant
under any type of\ Lorentz transformations [92]. The choice of a frame can
be interpreted as a choice of the reference system of an observer. It is
obvious that\ the observer's own dynamics, the state and structure of the
corresponding reference system, can affect the physical measurements,
including the determination of the energy and momentum of the gravitational
systems [93]. The frame dependence of \ these quantities is a physically
consistent feature, since the concepts that are valid in the special theory
of relativity are also valid in the general theory. The introduction of the
gravitational field does not change the frame dependence of the
gravitational energy-momentum energy. These quantities are not covariant
under local\ Lorentz transformations, but covariant only under global\
Lorentz transformations.

From a theoretical point of view, covariance under local Lorentz
transformations is a desirable feature for any physical theory, but, at the
experimental level, this symmetry breaking would not be detectable, as the
metric tensor remains unchanged regardless of the privileged orientation of
reference frames [94].\

\subsection{ Issue of Lorentz invariance in TG}

In order to make an equation generally covariant, a connection\ is always
necessary as is well known in analytical mechanics and gauge theories.\ A
general coordinate transformation\ introduces an affine connection, while a
local Lorentz transformation introduces an inertial connection.

The connection is not a tensor, and it's vanishing is not covariant. The
vanishing of the connection in the pure-tetrad formulation breaks the local
Lorentz invariance, which is often erroneously attributed to teleparallel
gravity itself. This confusion is due to a misunderstand which ignores a
fact i.e. the pure tetrad TG is written in the specific class of frames.
After fixing the class of frames, the theory is no longer manifestly local
Lorentz invariant. The question of local Lorentz invariance is\textbf{\ }%
ill-defined in this case since we choose the specific frame and hence we are
not allowed to perform local Lorentz transformations. The whole discussion
of local Lorentz invariance in this case seems rather misguided and not an
indicator of any problem of TG. The analogous situation in electromagnetism
is the discussing of the gauge invariance in a specific class of gauge (the
Coulomb gauge or the Landau gauge, for example), which obviously does not
make sense [27, 29]. On the other hand, provided the Lorentz connection is
appropriately taken into account, TG must be fully invariant under local
Lorentz transformations [27]. In particular, the question of local Lorentz
invariance has been resolved due to the existence of the inertial Lorentz
connection [95].

The physical relevance of\ local Lorentz covariance is just to ensure the
theory being\ valid in the frame of any observer in space-time. This is the
main feature of the local Lorentz covariance and an issue of consistency of
the theory [95].

\subsection{Local Lorentz group is not a genuine gauge group}

The role played by the Lorentz connection seems to be a constant source of
serious debates\ in the literature. Except in GR, whose Lorentz connection
includes both gravitation and inertial effects, in all other relativistic
theories the Lorentz connection has to do with inertial effects only.
Especially, it is alien objects and\ have nothing to do with gravitation. It
describes only the inertial effects present in a given class of frames.

The variation of the action with respect to the Lorentz connection $%
A^{a}{}_{b\mu }$ is a\ surface term and vanishes identically given
appropriate boundary conditions [29.] This variation does not produce
additional equations of motion. The field equations of neither TG nor GR are
able to determine the Lorentz connection. The local Lorentz group is
obviously a kinematic and not a dynamic symmetry. One should not expect,
therefore, any dynamic effect coming from a gaugefication\ of the Lorentz
group. Therefore, as far as the solutions of the field equations are
concerned, the Lorentz connection can be chosen arbitrarily in order to
solve the field equations. In particular, this allows us to set it to\ zero
and effectively obtain the purely tetrad TG. However, the crucial point is
that the Lorentz connection can be chosen arbitrarily only when we are
interested in solutions of the field equations. In general cases, the
Lorentz connection still plays an important role as it contributes to the
action through the surface term. In many situations the total value of the
action is important, for example, the calculations of the energy-momentum
and black hole thermodynamics.

Although they produce physical effects and have energy and momentum, as
inertial effects Lorentz connections cannot be interpreted as a field in the
usual sense of classical field theory. They\ have no sources and can be
generated canceled by coordinate transformations. Their action in general do
not vanish at infinity [96]. Moreover, Local Lorentz transformations are not
generated by first class constraints, as is usual in ordinary gauge theories
[95]. In other words, there is no physical gauge field corresponding to the
local Lorentz group [10].

\subsection{Genuine Gravity}

An important issue in any gravity theory is the choice of the dynamic
variables. In formulating GR, Einstein chose the metric $g_{\mu \nu }$\ as
the fundamental object to describe the gravitational potential. The
connection $\Gamma ^{\alpha }{}_{\mu \nu }$ $=\left\{ _{\mu }{}^{\alpha
}{}_{\nu }\right\} _{g}$ is assumed to represent the gravitational field
strength. It is not a tensor and can be switched off and set to zero at
least in a point according to the equivalent principle. Both $g_{\mu \nu }$
and $\Gamma ^{\alpha }{}_{\mu \nu }$ include the gravitation $B^{a}{}_{\mu }$
as well as inertial effects $A^{a}{}{}_{b\mu }$.

In TG the gravitational field is described by the torsion tensor $T^{\lambda
}{}{}_{\mu \nu }$, the fundamental quantity is the frame field $h^{a}{}_{%
\mu
}$, while\ the connection $A^{a}{}{}_{b\mu }$ only reflects the inertial
properties of the frame [29], at least in most cases. In the purely tetrad
TG\ $A^{a}{}{}_{b\mu }$ is set to\ zero and only\ $B^{a}{}_{\mu }$\ is left
representing gravity [97].

In STG, $R^{a}{}_{b\mu \nu }=0$, $T^{a}{}_{\mu \nu }=0$, one can set $%
A^{a}{}{}_{b\mu }=0$ and $\Gamma ^{\lambda }{}_{\mu \nu }$, but the
gravitational field $F^{a}{}_{\nu \mu }=\partial _{\nu }B^{a}{}_{\mu
}-\partial _{\mu }B^{a}{}_{\nu }\neq 0$.

We see that among all variables in the three approaches of gravity theories,
only the translational \emph{connection} $B^{a}{}_{\mu }$ can not be assumed
being zero. This means that only $B^{a}{}_{\mu }$ represent the \textbf{%
genuine gravity}. However, it does not appear in the theories explicitly,
but hides in them in terms of $h^{a}{}_{\mu }$ and $g_{\mu \nu }$. The
essential nature of gravity is a translation gauge field $B^{a}{}_{\mu }$.
However, its theoretical formulations usually take a implicit appearance. GR
or STG is its Newton's type, while TG is its Yang-Mills-like formulation.
Lorentz group and connection have nothing to do with gravity, therefore in
the metric formulations of gravity theory the explicit appearance of the
Lorentz connection is avoided.

\section{Conclusion}

Einstein`s great contribution is introduction of the concept of transformations (special relativity) and the geometrization of gravity (GR).
Transformation theory is the essence of the new method of theoretical physics [98]. According to Klein's Erlanger Program, geometry deals with the
invariant properties under transformation groups. The study object of modern differential geometry is manifold. In physics, spacetime is a four-dimension
manifold and gravity is equivalent to the geometry of the spacetime according to Einstein. The most original geometrical structures on a
manifold are coordinate systems and tangent vectors. The set of the tangent vectors at a point on the manifold constitute a linear space, the tangent
space. Then a tangent bundle structure is set up on the manifold which is the base space. There exist two kinds of transformations: coordinate
transformations on the base space and the linear transformations in the tangent space. At the same time, the tangent space is isomorphic to a\textbf{
\ point set} $\mathbb{R}_{4}$ and then there exists another kind of transformations, i.e. translations. Consequently, the spacetime is endowed
with the structure of principal fiber bundles, of which one is the frame bundle and another is the translation bundle. The structure group of the
frame bundle consists of the base transformations of the tangent space including the transformations of the coordinate bases, while the structure
group of the translation bundle consists of the translations in $\mathbb{R}
_{4}$. On every principal fiber bundle there exists the corresponding connection, i.e. the affine connection $\Gamma ^{\lambda }{}_{\mu \nu }$ and
the translation connection $B^{a}{}_{\mu }$.

In addition to vector fields, tensor and differential form fields also can
be defined on the spacetime manifold. Especially, physics endows the
spacetime with some metric structure, for example, Lorentz metric. Whenever
a Lorentz metric is defined on the spacetime, a subbundle is constructed.
Its structure group is the Lorentz group $SO(3,1)$, the fiber consists of
Lonrentz orthonormal frames (tetrads) and the corresponding connection is
the Lorentz connection $A^{a}{}_{b\mu }$.

The basic geometrical objects on the spacetime manifold are the sets of the
basis vectors, the frames which solder the frame bundle with the associated
tangent bundle. As the most basic geometrical objects, the frames can be
classified and endowed with different physical meaning according to their
geometrical characters. There exists a kind of frames $h^{a{}}{}_{\mu
}:=\partial _{\mu }x^{a{}}{}+A^{a}{}_{b\mu }x^{b}+B^{a{}}{}_{\mu }$
corresponding to the geometry structure above mentioned. A holonomic frame $%
e^{b{}}{}_{\mu }=\partial _{\mu }x^{b}$, for example, describes a \emph{%
inertial reference system. }The anholonomic frame $h^{\left( 0\right)
b{}}{}_{\mu }=\partial _{\mu }x^{b}+B^{b{}}{}_{\mu }$ represents a proper
reference system in the gravitational field, while the anholonomic frame $%
h^{\left( L\right) b{}}{}_{\mu }=\partial _{\mu }x^{b}+A^{b}{}_{c\mu }x^{c}$
represents a\emph{\ }non-inertial reference system. The translation
connection $B^{b{}}{}_{\mu }$ is the gravitational potential. However, the
Lorentz connection $A^{b}{}_{c\mu }$ plays the role of the inertial force
appearing in the\emph{\ }non-inertial reference system, the term $%
A^{b}{}_{c\mu }x^{c}$ plays the role of the inertial "potential".

Starting from this geometry structure three equivalent gravity theories, GR,
TG and STG can be constructed. The difference between them is the choice of
dynamic variables.

In GR and STG, $g_{\mu \nu }$\ represents the gravitational potential and $%
\Gamma ^{\rho }{}_{\mu \nu }=\left\{ _{\mu \nu }^{\rho }\right\} $ the
gravitational field strength. $R^{a}{}_{b\mu \nu }$ is not dynamic and then
is dropped out in STG. In all the three theories $R^{a}{}_{b\mu \nu }$ has
nothing to do with gravity. The statement "gravity is the curved spacetime"
is a misinterpretation. The the essence of gravity is the translation
connection $B^{a}{}_{\mu }$ and its curvature i. e. the torsion $T^{\rho
}{}_{\mu \nu }$. Lorentz group and its connection do only with inertial and
then can be avoided in the metric formulations of gravitational theories.
Although the essence of gravity is $B^{a}{}_{\mu }$, but the best
representation of it is $g_{\mu \nu }$. Einstein's genius is the choice of
the variable $g_{\mu \nu }$ and the identifying gravity as geometry. \

STG brings a new perspective to bear on GR.\ It can uncover the more
fundamental nature of the gravitational interaction. It inherits Einstein's
great heritages, the $\Gamma \Gamma $ Lagrangian, the field equation and the
gravitational energy-momentum expression, but abandons the redundant baggage
i.e. curvature. It might be closer to Einstein's own view of GR than what
has become conventionally established as \textquotedblleft Einstein's
GR\textquotedblright .\ STG fundamentally \textbf{deprives} gravity of any
inertial character and realizes a purified gravity theory [19].\

Although GR, TG and STG are equivalent each other, their extensions $f\left(
R\right) $, $f\left( T\right) $ and $f\left( Q\right) $ are not. For the
accelerated expansion of the universe they all seek a geometrical origin and
give different explanation. The true meanind of $f\left( Q\right) $ is that
the acceleration of the universe expansion is just the effect of the
translation field $B^{a}{}_{\mu }$ rather than some mysterious dark energy.
In other words, the accelerated expansion of the universe is the intrinsic
character of the spacetime geometry as the result of Einstein's spiritual
heritage. However, at the present stage of the research, there is no final
probe discriminating between $f\left( R\right) $, $f\left( T\right) $ and $%
f\left( Q\right) $ theories. Furthermore, the bulk of observations to be
considered is very large and then an effective Lagrangian addressing the
whole phenomenology at all astrophysical and cosmic scales, would be very
difficult to find.

\textbf{\ Acknowledgments }
 The research work is supported by   the National Natural Science Foundation of China (11645003), and supported by  LiaoNing Revitalization Talents Program (XLYC2007047).

\end{document}